# Nitrogen based electride superconductor $Nb_5Ir_3N$ under pressure: multifunctional physical properties from DFT based first-principles investigation

M. Abdul Hadi Shah[1,2], J.H. Abir[2], S.H. Naqib[2,*]

[1]Department of Physics, Rajshahi University of Engineering and Technology, Rajshahi 6204, Bangladesh
[2]Department of Physics, University of Rajshahi, Rajshahi 6205, Bangladesh
*Corresponding author; Email: salehnaqib@yahoo.com (S.H. Naqib)

## Abstract

Discovery and study of superconducting electrides have opened new avenues in condensed matter physics, driven by their intriguing multifunctional features spanning ambient and high-pressure regimes. This work investigates the ternary nitride superconductor $Nb_5Ir_3N$ under pressure ranging from 0 to 20 GPa via density functional theory based simulations. Estimated structural parameters agree well with the available data, confirming their reliability and supporting the validity of this analysis. Well-converged Birch–Murnaghan equations of state and negative formation enthalpy, together with evaluated elastic constants and phonon spectra, confirm the structural, thermodynamic, mechanical, and dynamical stability of $Nb_5Ir_3N$ over the entire pressure range. Pressure dependent elastic constants and polycrystalline elastic moduli are investigated. The compound is categorized as ductile in light of estimated mechanical indices. Elastic anisotropy factors indicate that $Nb_5Ir_3N$ remains anisotropic, with the degree of anisotropy gradually decreasing as pressure increases. Electronic band structure and density of states of $Nb_5Ir_3N$ are calculated with and without spin–orbit coupling to investigate its influence on the electronic structure. Calculated optical response reveals substantial intraband contributions in the low-energy region, intense ultraviolet absorption, and strong reflectivity. The spectra exhibit optical anisotropy and a broadening at higher pressures. Pressure-induced superconducting features are also discussed qualitatively.

## 1. Introduction

Electride is a kind of novel/stoichiometric ionic solid in which electrons serve as the anions [1–4]. Electrides have gained significant interest in both fundamental research and technological applications due to their intriguing properties, including antiferromagnetism [1], ultralow work functions [5], high electronic mobility [6], exceptional catalytic activity [7], and distinct anisotropic electronic and optical features [3,8]. Since the discovery of the first crystalline organic electride in 1983 [9] and room-temperature inorganic electride in 2003 [10], a number of organic and inorganic electrides have been successfully synthesized, and their unique electronic structures are evolved through both computational and experimental routes [11–17]. Due to the difficulties during the synthesis route of organic electrides [13], condensed matter physicists are now decisively focusing on inorganic electrides [8,18].

Superconductivity in electrides was first observed in mayenite, which has a cage-like structure with a superconducting transition temperature, $T_c$ of 0.4 K [19]. Since then, substantial research groups focused on investigating superconductor–electride dualism over the decades, as the coexistence of superconducting and electride states remains relatively unexplored. In electrides, a fraction of electrons detaches from atoms and localizes within interstitial voids, whose topology can be tuned to achieve novel properties [14]. Although alkali metal electrides under high pressure exhibit moderate superconductivity ($T_c$ < 20 K) [20,21], the incorporation of nonmetals into electrides can modify the behavior of interstitial electrons (IEs), thereby enhancing superconducting performance with high $T_c$ > 40 K [22,23]. It is noteworthy that binary compounds constitute a significant proportion of the predicted superconducting electrides reported to date [3,5,24–27], compared with ternary compounds [18,28,29].

Among three types of $A_5B_3$-type binary compounds, $Cr_5B_3$-type [30] and $W_5Si_3$-type [31,32] belong to tetragonal symmetry, while $Mn_5Si_3$-type structure consists of hexagonal crystal symmetry [33–36]. An $A_5B_3$-type binary material of any of the above types may exhibit superconducting properties, or it may not. Nonetheless, some Nb-based binary systems have been rigorously studied for their superconducting properties across various crystal symmetries. For instance, Koch *et al*. synthesized a stable tetragonal phase of $Nb_5Ir_3$ with a critical temperature ($T_c$) of 2.8 K [37]. In contrast, Zhang *et al*. investigated $Nb_5Ir_3$ in both tetragonal and hexagonal symmetries while introducing an electride state, reporting a $T_c$ of 9.8 K [38].

Incorporating an interstitial Z atom at the center of the $Mn_6$ octahedra within a hexagonal $Mn_5Si_3$-type structure transforms it into a ternary system without altering its symmetry, which will be the key challenge [18]. This modification could lead to significant advancements in condensed matter physics, particularly in the study of superconductivity. The inclusion of interstitial atoms (typically O or C) often acts as electron donors or acceptors, which affects the electronic properties of the material. Therefore, these changes significantly enhance the physical properties of material, such as its resistance to high-temperature corrosion [39], its superconductivity [38,40], and its potential as a highly effective anode material for ion batteries [41,42]. In particular, in the Nb-based system, inclusion of small atoms (such as C or O) may induce or significantly enhance $T_c$ [38,43]. For instance, the tetragonal phase of binary $Nb_5Ir_3$ is converted into hexagonal ternary $Nb_5Ir_3O_x$ (x = 1.0) when substituting interstitial O atom, raising the $T_c$ from 9.8 K to 10.5 K [38]. As a consequence of growing interest, Nb-based other systems $Nb_5(Ge/Pt)_3O_x$ (x = 1.0) have also been investigated [43,44].

Remarkably, Nitrogen (N) atom in interstitial positions within filled $Mn_5Si_3$-type compounds are exceedingly scarce. Nonetheless, several studies have successfully explored compounds with Nitrogen (N) atom in interstitial positions such as $V_5Si_3N$ [45], $Zr_5Sn_3N$ [46], $Ti_5Si_3N$ [47], and $La_5Ge_3N$ [48], all of which unequivocally lack superconducting properties. Moreover, N-atoms in superconductors play a pivotal role in strengthening chemical bonding and enhancing electron-phonon coupling, which collectively drives the superconducting nature [49,50]. To address, recently, Yang *et al*. proposed a synthesis route of N-atom-filled ternary superconductor $Nb_5Ir_3N$ [18]. They reported not only a synthesis strategy but also probes the superconducting features ($T_c$ = 8.7 K) along with electronic features considering spin-orbit coupling (SOC) at ambient (0 GPa) pressure.

The investigation of physical properties is strongly influenced by external parameters, including temperature, pressure, epitaxial strain, and electric or magnetic fields. Among these parameters, pressure plays a crucial role as a fundamental thermodynamic variable that governs the properties of materials. It can induce significant modifications in the microscopic structure, interatomic electrostatic interactions, electronic orbital configurations, and chemical bonding characteristics. Therefore, induction of pressure on a compound through experimental roots is a demanding venture. Pressure also serves as a crucial parameter for exploring variations in ground-state physical properties through theoretical approach within the first-principles DFT computations [51]. Such approaches are particularly valuable for systems exhibiting unconventional electronic features, topological order, and enhanced superconductivity [3,52,53]. Consequently, pressure has been widely used as a powerful tool for decades in the discovery of materials inaccessible at ambient conditions [54].

Despite the reported superconducting features of $Nb_5Ir_3N$ at ambient pressure, a comprehensive understanding of its structural, phonon, mechanical, hardness, thermophysical, anisotropic, electronic, and optical properties is still lacking. Furthermore, the effects of pressure on these physical properties have not yet been systematically investigated. To address these gaps, the present study examines the fundamental physical properties of $Nb_5Ir_3N$ from ambient pressure up to 20 GPa. The obtained insights is anticipated to support future experimental studies and evaluate the prospects of $Nb_5Ir_3N$ as a promising multifunctional material for high-pressure applications.

## 2. Computational approach

In this study, a plane-wave pseudopotential approach is employed using DFT simulations [55] integrated within the CASTEP package in Materials Studio 2020 [56] to investigate the structural and electronic features of $Nb_5Ir_3N$ electride. The Perdew-Burke-Ernzerhof with solids correction (PBEsol) in the generalized gradient approximation (GGA) [57] as the exchange-correlation functional is chosen. An ultrasoft pseudopotential of Vanderbilt-type is adopted for the interaction of charges between atomic cores and valence electrons. The solids-corrected version of the GGA approximation introduced by Perdew *et al.* for densely packed solids provides a more precise physical nature than the conventional version [58]. We implemented a plane-wave basis set cutoff of 350 eV and a Monkhorst-Pack k-grid of 10×10×7 (comprising 56 irreducible *k*-points), for Brillouin zone (BZ) integrations. A popular optimization algorithm of Limited-memory Broyden–Fletcher–Goldfarb–Shanno (LBFGS)-type [59] is utilized to optimize equilibrium structural parameter. Additionally, the Pulay density mixing scheme was chosen with a charge mixing amplitude of 0.5 and a charge mixing cut-off of 1.5 for the electronic minimizer. Spin–orbit coupling (SOC) effects were also considered in the electronic structure calculations. Accordingly, computations of band structure and density of states were performed both with and without SOC using QUANTUM ESPRESSO code [60].

Phonon dispersion curves (PDC) and phonon density of states (PHDOS) are calculated within the density functional perturbation theory (DFPT) based on the finite displacement supercell method [61,62] within the CASTEP package. A cutoff radius of 2.0 Å were employed during phonon calculations keeping energy cut-off and *k*-grid similar for all pressures.

The elastic constants of materials can be estimated by the stress-strain method [63] implemented within the Hooke's law:

$$\sigma_{ij} = C_{ijkl}\varepsilon_{kl}, \tag{1}$$

where $\sigma_{ij}$ is the stress tensor, $C_{ijkl}$ is the elastic constant tensor which is a 6×6 matrix, and $\varepsilon_{kl}$ represents the Lagrangian strain tensor.

The symmetry of a crystal structure determines the form of its elastic stiffness tensor, causing some tensor components to be equal by symmetry while others become identically zero. In the most general case, an anisotropic crystal possesses 21 independent elastic stiffness constants ($C_{ij}$). However, owing to the high structural symmetry of the hexagonal crystal system, the elastic stiffness matrix is reduced to only five independent elastic constants: $C_{11}$, $C_{12}$, $C_{13}$, $C_{33}$ and $C_{44}$ [64]. The remaining elastic constants are either related by symmetry [e.g., $C_{66} = (C_{11}- C_{12})/2$] or zero.

The calculated single-crystal elastic constants can be used to estimate the polycrystalline elastic properties, including the bulk modulus ($B$) and shear modulus ($G$), using the Voigt–Reuss–Hill (VRH) approximation [65–67]. Within this framework, the effective elastic moduli ($B$ and $G$) are obtained by averaging the Voigt and Reuss bounds and can be expressed in terms of the independent elastic constants as follows [68]:

$$\left.\begin{aligned}
G &= \frac{G_V + G_R}{2};\ B = \frac{B_V + B_R}{2},\ \text{where,}\\
G_V &= \frac{1}{30}(C_{11} + C_{12} + 2C_{33} - 4C_{13} + 12C_{44} + 12C_{66}),\\
G_R &= = \frac{5[(C_{11} + C_{12})C_{33} - 2C_{13}^2]C_{44}C_{66}}{6B_V C_{44}C_{66} + 2[(C_{11} + C_{12})C_{33} - 2C_{13}^2](C_{44} + C_{66})},\\
B_V &= \frac{1}{9}[2(C_{11} + C_{12}) + 4C_{13} + C_{33}],\\
B_R &= \frac{(C_{11} + C_{12})C_{33} - 2C_{13}^2}{C_{11} + C_{12} + 2C_{33} - 4C_{13}},
\end{aligned}\right\} \tag{2}$$

where $B_V$ ($B_R$) and $G_V$ ($G_R$) denote the upper (lower) limit for polycrystalline solid at the Voigt (Reuss) boundary.

The polycrystalline Young's modulus ($Y$) and Poisson's ratio ($\sigma$) can be calculated as [69]:

$$Y = \frac{9BG}{3B + G}, \sigma = \frac{3B - 2G}{2(3B + G)}. \tag{3}$$

The optical properties of materials are characterized by their dielectric function $\varepsilon(\omega)$ which is a complex tensor that explains the linear response of electronic system to the electromagnetic radiation, and expressed as:

$$\varepsilon(\omega) = \varepsilon_1(\omega) + i\varepsilon_2(\omega), \tag{4}$$

where $\varepsilon_1(\omega)$ and $\varepsilon_2(\omega)$ correspond to the real and imaginary parts of the dielectric constants, respectively.

The frequency-dependent imaginary dielectric function $\varepsilon_2(\omega)$ indicate the absorption of the incident radiations and is expressed by [70]:

$$\varepsilon_2(\omega) = \left(\frac{e^2\hbar}{\pi m^2\omega^2}\right)\sum_{v,c}\int_{BZ} |M_{cv}(k)|^2 \delta[\omega_{cv}(k) - \omega]d^3k, \tag{5}$$

where $M_{cv}(k) = \langle u_{ck}|\delta\nabla|u_{vk}\rangle$ describes the momentum dipole matrix components for direct transitions between valence $u_{vk}(r)$ and conduction band $u_{ck}(r)$ electronic states at the wave vector $k$, $\nabla$ is the momentum operator, and $\hbar\omega_{cv}(k) = (E_{ck} - E_{vk})$ corresponds to the transition energy. The integral is taken over the first BZ.

The real dielectric constant $\varepsilon_1(\omega)$ can be derived from the imaginary part $\varepsilon_2(\omega)$ using Kramers-Kronig relation as [71]:

$$\varepsilon_1(\omega) = 1 + \frac{2}{\pi}P\int_0^\infty \frac{\omega'\varepsilon_2(\omega')}{\omega'^2 - \omega^2}d\omega', \tag{6}$$

where $P$ signifies the principal value of the integral.

To acquire deeper understanding of optical nature, the complex refractive index N($\omega$) is to be estimated. Being a complex quantity, it has two parts: the real part, referred to as refractive index $n(\omega)$ and the imaginary part, known as the extinction coefficient $k(\omega)$. Therefore, N($\omega$) [= $n(\omega)$ + i$k(\omega)$] is calculated using the following expressions [72]:

$$\left.\begin{aligned} n(\omega) &= \frac{1}{\sqrt{2}}\left[\sqrt{\varepsilon_1^2(\omega) + \varepsilon_2^2(\omega)} + \varepsilon_1(\omega)\right]^{1/2}, \\ k(\omega) &= \frac{1}{\sqrt{2}}\left[\sqrt{\varepsilon_1^2(\omega) + \varepsilon_2^2(\omega)} - \varepsilon_1(\omega)\right]^{1/2}. \end{aligned}\right\} \tag{7}$$

The other optical parameters *viz.* absorption coefficient $\alpha(\omega)$, optical conductivity $\sigma(\omega)$, reflectivity $R(\omega)$, and energy-loss function $L(\omega)$ are calculated using the expressions as introduced in literature [73,74].

The vibrational contributions to the thermodynamic properties are derived within the quasi-harmonic Debye model using the Gibbs2 code [75] to compute temperature-dependent specific heat capacity at constant volume ($C_V$), vibrational internal energy (or enthalpy) ($U_V$), vibrational entropy ($S_V$), and vibrational Helmholtz free energy ($F_V$). Within this framework, thermal functions are derived from established thermodynamic relations, enabling a comprehensive evaluation of structural stability and performance across a broad thermal regime. The vibrational Helmholtz free energy is expressed as [76]:

$$F_V[\Theta_D(V);T] = nk_BT\left[\frac{9\Theta_D}{8T} + 3ln\left(1 - e^{-\frac{\Theta_D}{T}}\right) - D\left(\frac{\Theta_D}{T}\right)\right], \tag{8}$$

where $n$ is the number of atoms per formula unit, $D\left(\frac{\Theta_D}{T}\right)$ represents the Debye integral [75].

Debye temperatures $\Theta_D$(V) are calculated by using quasi-harmonic Debye model from the static bulk moduli $B_S$(V) and the Poisson's ratio ($\sigma$) [77]. In this approach, $\Theta_D$ can be expressed as: $\Theta_D = (\hbar/k_B)\left[6\pi V^{1/2}n\right]^{1/3} f(\sigma)\left(\frac{B_s}{M}\right)^{1/2}$, where $M$ is the molecular mass.

The vibrational entropy and the specific heat capacity at constant volume is represented as [78]:

$$S_V = nk_B\left[4D\left(\frac{\Theta_D}{T}\right) - 3ln\left(1 - e^{-\Theta_D/T}\right)\right], \tag{9}$$

$$C_V = 3nk_B\left[4D\left(\frac{\Theta_D}{T}\right) - \frac{3\Theta_D/T}{e^{\Theta_D/T} - 1}\right], \tag{10}$$

where $\alpha = \frac{\gamma C_V}{B_T V}$, $C_P = C_V(1 + \alpha\gamma T)$; $\gamma = \frac{d\ ln\Theta_D(V)}{d\ lnV}$, $B_T$ and $\gamma$ are known as isothermal bulk modulus and Grüneisen parameter, respectively.

## 3. Results and discussion

### *3.1. Structure and stability*

#### *3.1.1. Structural stability analysis*

$Nb_5Ir_3N$ belongs to hexagonal structure with space group $P6_3/mcm$ (#193) which contains two formula unit. In this structure, Nb atom occupies two positions; Nb1 atoms in elementary cell are occupied at Nb1 (4*d*): (0.000, 0.3333, 0.6667), while Nb2 (6*g*) at (0.2348, 0.2348, 0.75) positions. In contrast, Ir (6*g*) and N (2*b*) are positioned at (0.5887, 0.5887, 0.75) and (0.000, 0.000, 0.000) Wyckoff sites, respectively [18]. **Figure 1a-b** depicts the unit cell structure of $Nb_5Ir_3N$ in both two dimensions (2D: **Fig. 1a**) and three dimensions (3D: **Fig. 1b**), constructed using the VESTA software.

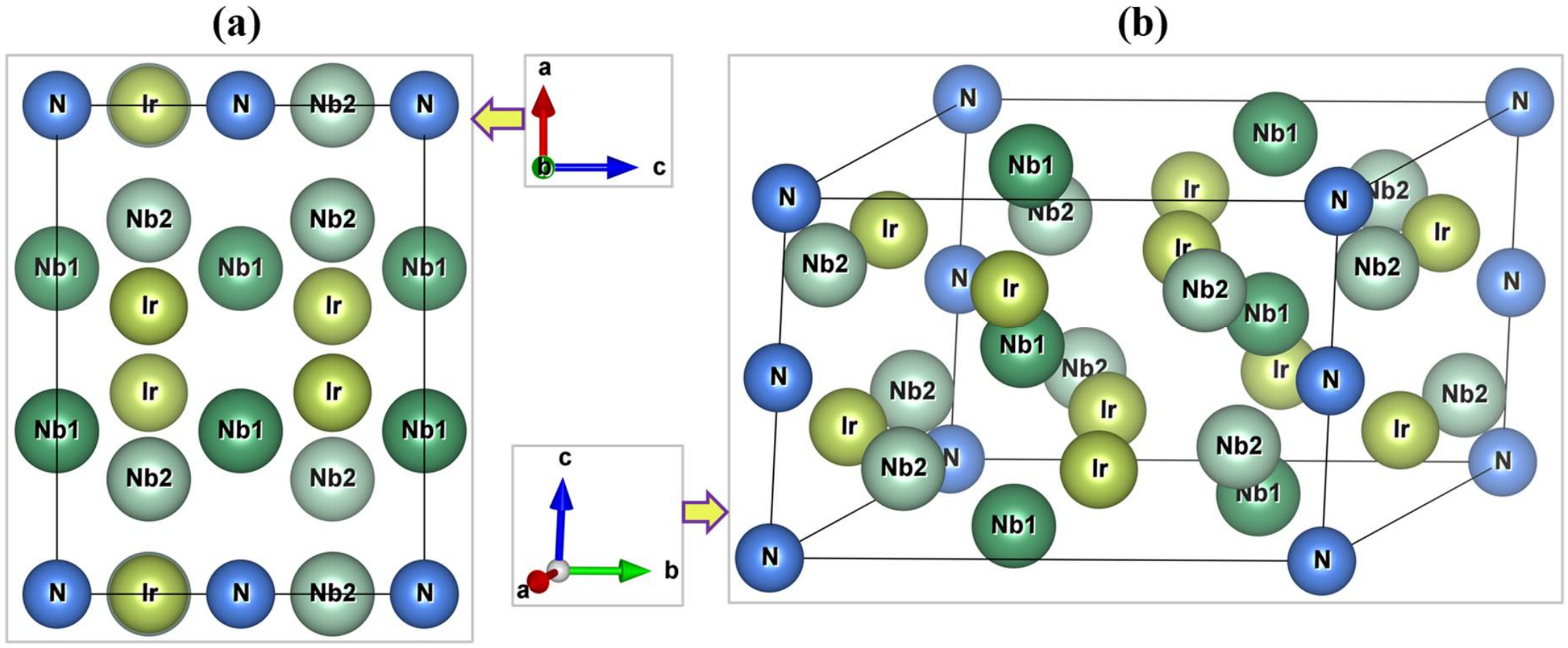

**Fig. 1**. Unit cell structure of $Nb_5Ir_3N$ in real space: (a) 2D view and (b) 3D view.

In order to achieve equilibrium structural parameters at ambient (0 GPa) pressure, we have initially considered various functionals (such as, LDA, GGA-PBE, GGA-RPBE, GGA-PBESol, GGA-PW91) for volume optimization (not given here). Among the available functionals, GGA-PBEsol provides the least deviation of volume at approximately 0.31% in respect to the experimental data [18]. Accordingly, this functional is adopted to examine the physical properties of $Nb_5Ir_3N$ under ambient and elevated pressures. For convergence testing, a *k*-point mesh was fixed initially at (10×10×7) and then the plane-wave cut-off energy was systematically increased from 180 eV. A converged cut-off energy of 350 eV was identified from the point where the total energy curve flattened out. Subsequently, using this optimized cut-off energy (350 eV), *k*-point convergence was then assessed by progressively refining the mesh density from 2×2×1 while keeping the cut-off fixed. Based on the results of these two convergence tests, the final plane-wave cut-off energy of 350 eV and *k*-point mesh of (10×10×7) were used for all subsequent calculations.

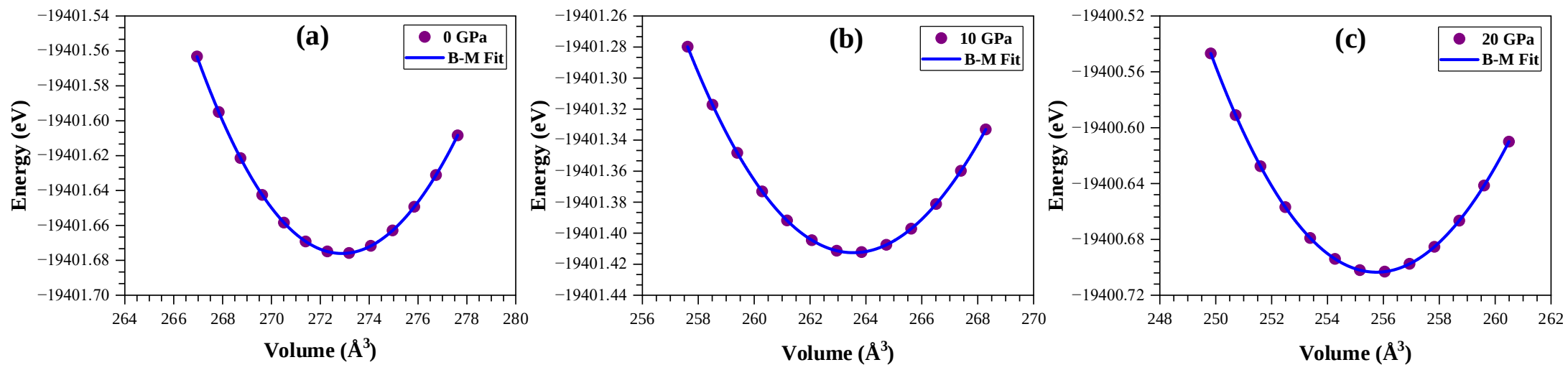

**Fig. 2**. Variation of the total energy with unit cell volume obtained using the GGA–PBESol functional of $Nb_5Ir_3N$ for (a) 0 GPa, (b) 10 GPa, and (c) 20 GPa.

The energy minimization process is used to regulate the structural properties of compounds. The calculated total energies (*E*) with respect to unit cell volume (*V*) for $Nb_5Ir_3N$ at 0 GPa, 10 GPa, and 20 GPa are plotted in **Fig. 2**. To obtain the equilibrium lattice parameters at $P = 0$ GPa and $T = 0$ K, *E-V* data were fitted to the Birch-Murnaghan (B-M) equation of state (*EOS*) [79,80]:

$$E(V) = E_0 + \frac{9V_0B_0}{16}\left\{\left[\left(\frac{V_0}{V}\right)^{\frac{2}{3}} - 1\right]B_0' + \left[\left(\frac{V_0}{V}\right)^{\frac{2}{3}} - 1\right]^2\left[6 - 4\left(\frac{V_0}{V}\right)^{\frac{2}{3}}\right]\right\}. \quad (11)$$

The calculated optimized structural parameters of $Nb_5Ir_3N$ are listed in **Table 1**. The calculated equilibrium lattice parameters and volume decreases with increasing pressure emphasizing that $Nb_5Ir_3N$ undergoes progressive compression under applied pressure. This behavior is expected for a mechanically stable solid and reflects the reduction of interatomic distances as the external pressure increases. Relative to the ambient-pressure, a maximum volume reduction of ~6.70% up to 20 GPa indicates that $Nb_5Ir_3N$ is compressible in keeping its structural integrity, assuming no discontinuities appear in the equation of state (EoS) or other estimated properties.

**Table 1.** Equilibrium lattice constant (*a*, *c* in Å), hexagonal ratio (*c/a*), volume ($V_0$ in Å$^3$), energy ($E_0$ in eV), bulk modulus ($B_0$ in GPa), pressure derivative of bulk modulus ($B_0'$), density ($\rho$ in gm/cm$^3$), formation energy ($E_f$ in eV/atom), and cohesive energy ($E_c$ in eV/atom) of $Nb_5Ir_3N$.

| $P$ | $a$ | $c$ | $c/a$ | $V_0$ | $E_0$ | $B_0$ | $B_0'$ | $\rho$ | $E_f$ | $E_c$ | Reference |
|---|---|---|---|---|---|---|---|---|---|---|---|
| 0 | 7.832 | 5.137 | 0.656 | 272.88 | -19401.682 | 270.6 | 4.50 | 12.84 | -1.272 | 11.263 | [This work] |
| 0 | 7.8398[a] | 5.1108[a] | 0.6519[a] | - | - | - | - | - | - | - | [18][a] |
| 0 | 7.86792[b] | 5.18481[b] | 0.6590[b] | - | - | - | - | - | - | - | [18][b] |
| 10 | 7.748 | 5.069 | 0.654 | 263.55 | -19401.418 | 307.2 | 4.83 | 13.30 | -1.276 | 11.249 | [This work] |
| 20 | 7.676 | 5.012 | 0.653 | 255.75 | -19400.709 | 351.4 | 4.98 | 13.70 | -1.285 | 11.209 | [This work] |

[a]$^{Expt.}$, [b]$^{Th.}$

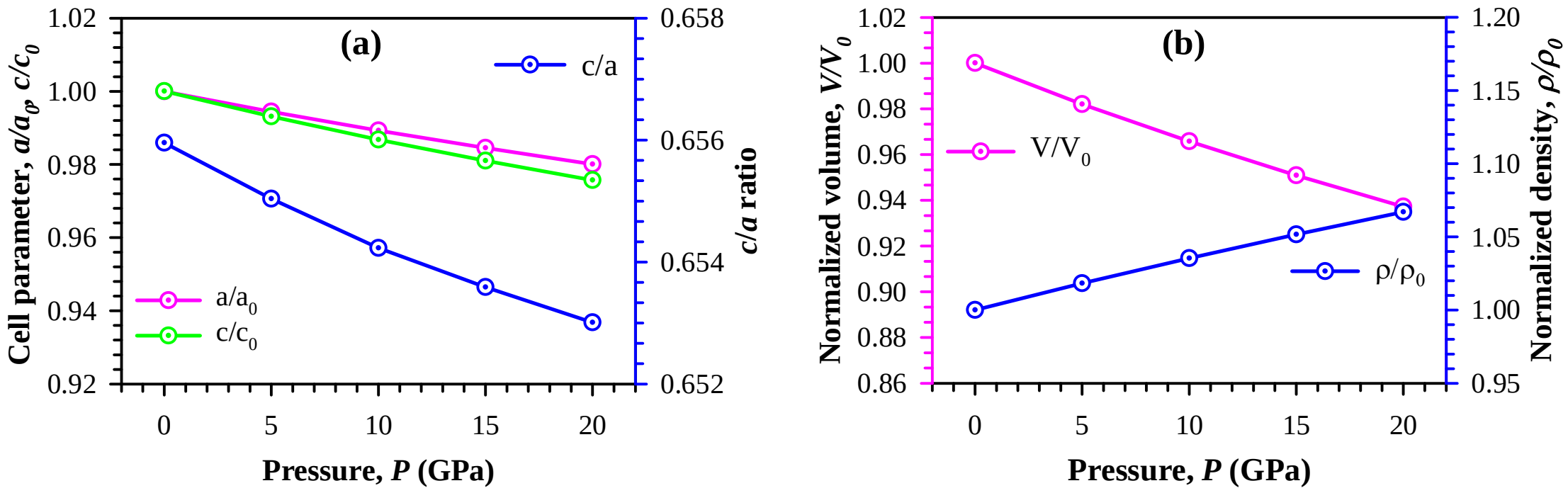

**Fig. 3**. Variation of (a) lattice parameters and *c*/*a* ratio, (b) normalized volume and density with hydrostatic pressure for $Nb_5Ir_3N$.

The pressure-induced variations in the normalized lattice parameters ($a/a_0$ and $c/c_0$), normalized density ($\rho/\rho_0$), and normalized unit-cell volume ($V/V_0$) of $Nb_5Ir_3N$ are presented in **Fig. 3**. As pressure increases, both lattice parameters decrease monotonically (**Fig. 3a**), whereas the density increases accompanied by a continuous reduction in the unit-cell volume (**Fig. 3b**). These trends are in good agreement with the results reported by Zhang *et al.* [**81**]. The slightly more rapid decrease of $c/c_0$ compared with $a/a_0$ exhibits that *a*-direction is more resistant to contraction than *c*-direction in $Nb_5Ir_3N$, reflecting small anisotropic compressibility. Such anisotropic response originates from the directional nature and heterogeneous bond strength within the crystal structure, making the lattice more susceptible to compression along the *c*-axis under pressure.

### *3.1.2. Thermodynamical stability*

In order to validate the chemical phase stability condition of $Nb_5Ir_3N$, its formation energy ($E_f$) and cohesive energy ($E_c$) are to be examined. The formation energy forecasts the stability of crystals in respect of decomposition into its bulk constituent elements. Either the energy needed to form a crystal from free atoms or the work needed to decompose a compound into isolated atoms is described as the cohesive energy, which correlates with the structural stability. They are estimated within the relation [**82**]:

$$E_f = \frac{1}{9}\left(E^{total}_{Nb_5Ir_3N} - 5E^{bulk}_{Nb} - 3E^{bulk}_{Ir} - E^{bulk}_{N}\right), E_c = \frac{1}{9}\left(5E^{iso}_{Nb} + 3E^{iso}_{Ir} + E^{iso}_{N} - E^{total}_{Nb_5Ir_3N}\right), \quad (12)$$

where $E^{total}_{Nb_5Ir_3N}$ denotes the total energy of $Nb_5Ir_3N$, and $E^{\mathrm{bulk}}_{\mathrm{Nb}}$, $E^{bulk}_{Ir}$, and $E^{\mathrm{bulk}}_{\mathrm{N}}$ are the total energies of the corresponding atom in the bulk form, whereas $E^{\mathrm{iso}}_{Nb}$, $E^{iso}_{Ir}$, and $E^{\mathrm{iso}}_{\mathrm{N}}$ represent the energies of the corresponding isolated atoms. Generally, a negative value of $E_f$ and a positive value of $E_c$ in a structure indicate its chemical stability [**83**].

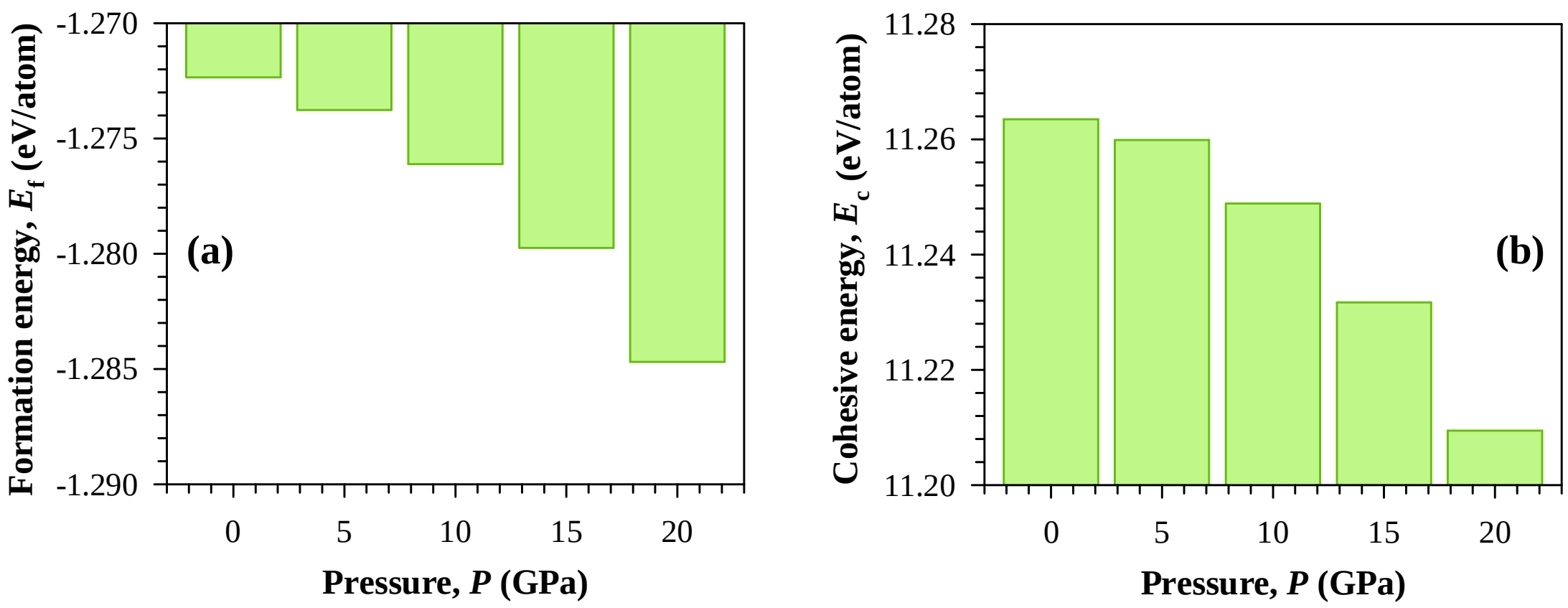

**Fig. 4**. (a) Formation energy and (b) cohesive energy as a function of pressure within 0-20 GPa for $Nb_5Ir_3N$.

The computed values of $E_f$ and $E_c$ (cf. **Table 1**) are negative and positive, respectively, within the hydrostatic pressure 0-20 GPa implies that the studied $Nb_5Ir_3N$ in hexagonal symmetry are energetically feasible to synthesize from the thermodynamics point of view within the pressure range considered. A more negative $E_f$ value indicates enhanced thermodynamic stability at 20 GPa, that is, the stability is further strengthened with increasing pressure from ambient conditions (cf. **Table 1**).

### *3.1.3. Dynamical stability and vibrational spectroscopy*

Phonon dispersion is a key aspect of lattice dynamics revealing insights into the structural stability and vibrational effects governing thermodynamic behavior [**84**]. Evaluating the stability of high-pressure phases is therefore essential; accordingly, phonon dispersion relations and phonon density of states (DOS) were calculated. A system is considered dynamically stable when all phonon frequencies at different *k*-points are real and positive, whereas the presence of negative frequencies signifies instability [**85**]. In this study, the phonon dispersion curves were obtained using the finite displacement supercell approach within the first Brillouin zone, as implemented in the CASTEP code, over a hydrostatic pressure range of 0–20 GPa at an interval of 5 GPa. The resulting phonon dispersion curves of $Nb_5Ir_3N$ at 0 GPa, 10 GPa, and 20 GPa along high-symmetry directions are presented in **Fig. 5a-c**, where the right panel of each plot represents their corresponding phonon density of states (PHDOS).

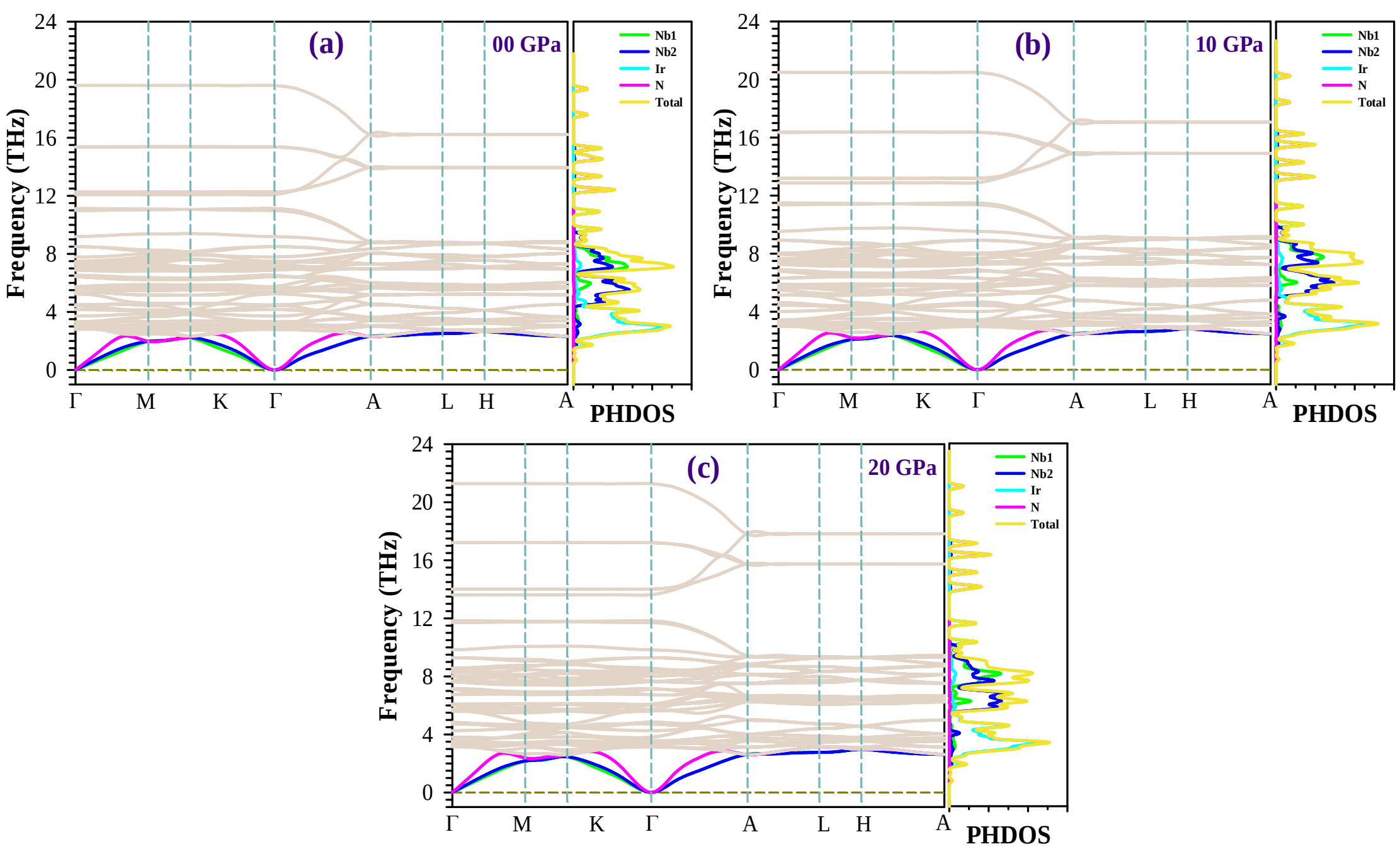

**Fig. 5**. Phonon dispersion spectra of $Nb_5Ir_3N$ at (a) 0 GPa, (b) 10 GPa, and (c) 20 GPa. The right panel represents their corresponding phonon density of states.

The absence of negative phonon frequencies at the *Γ*-point across the pressure range of 0 to 20 GPa (**Fig. 5a-c**) indicates the dynamical stability of $Nb_5Ir_3N$ throughout this pressure regime. This finding reinforces a robust cornerstone for understanding the evolution of the lattice dynamics and related physical properties under compression. The phonon spectra comprise three acoustic and fifty-one optical branches at all investigated pressures, reflecting the eighteen atoms in the formula unit. The overlap (or hybridization) between the acoustic and lower optical phonon branches in all the investigated pressures indicates the absence of a phononic band gap between these branches. This results in strong vibrational continuity and possible coupling between low- and high-frequency phonons, which can influence phonon scattering, lattice thermal transport, and electron–phonon interactions [**86**].

**Table 2**. Calculated phonon frequencies at zone-center (*Γ*-point) of infrared (*IR*)- and Raman (*R*)-active optical modes of $Nb_5Ir_3N$ electride at various pressures.

| Symmetry | Types (number) of modes | Frequency ($cm^{-1}$) at | | | Active modes for | | | | | |
|---|---|---|---|---|---|---|---|---|---|---|
| | | | | | Infrared (*IR*) at | | | Raman (*R*) at | | |
| | | 00 GPa | 10 GPa | 20 GPa | 00 GPa | 10 GPa | 20 GPa | 00 GPa | 10 GPa | 20 GPa |
| $E_{1u}$ | Acoustic (2) | 0 | 0 | 0 | × | × | × | × | × | × |
| $A_{2u}$ | Acoustic (1) | 0 | 0 | 0 | × | × | × | × | × | × |
| $E_{1u}$ | Optical (2) | 100.58 | 103.78 | 106.42 | √ | √ | √ | × | × | × |
| $E_{1u}$ | Optical (2) | 179.60 | 191.73 | 202.10 | √ | √ | √ | × | × | × |
| $E_{1u}$ | Optical (2) | 243.88 | 250.61 | 256.15 | √ | √ | √ | × | × | × |
| $E_{1u}$ | Optical (2) | 249.42 | 265.69 | 280.58 | √ | √ | √ | × | × | × |
| $E_{1u}$ | Optical (2) | 512.49 | 545.78 | 574.14 | √ | √ | √ | × | × | × |
| $A_{2u}$ | Optical (1) | 96.56 | 114.18 | 126.99 | √ | √ | √ | × | × | × |
| $A_{2u}$ | Optical (1) | 178.33 | 185.24 | 190.93 | √ | √ | √ | × | × | × |

| $A_{2u}$ | Optical (1) | 402.68 | 429.53 | 454.11 | √ | √ | √ | × | × | × |
|---|---|---|---|---|---|---|---|---|---|---|
| $E_{2g}$ | Optical (2) | 93.22 | 99.09 | 104.87 | × | × | × | √ | √ | √ |
| $E_{2g}$ | Optical (2) | 138.81 | 148.33 | 156.96 | × | × | × | √ | √ | √ |
| $E_{2g}$ | Optical (2) | 217.48 | 229.16 | 238.43 | × | × | × | √ | √ | √ |
| $E_{2g}$ | Optical (2) | 235.24 | 245.69 | 254.82 | × | × | × | √ | √ | √ |
| $E_{2g}$ | Optical (2) | 283.37 | 297.43 | 309.57 | × | × | × | √ | √ | √ |
| $A_{1g}$ | Optical (1) | 100.51 | 110.12 | 119.33 | × | × | × | √ | √ | √ |
| $A_{1g}$ | Optical (1) | 306.33 | 318.04 | 327.75 | × | × | × | √ | √ | √ |
| $E_{1g}$ | Optical (2) | 110.29 | 113.30 | 116.17 | × | × | × | √ | √ | √ |
| $E_{1g}$ | Optical (2) | 149.45 | 171.56 | 187.24 | × | × | × | √ | √ | √ |
| $E_{1g}$ | Optical (2) | 242.51 | 255.51 | 267.13 | × | × | × | √ | √ | √ |
| $E_{2u}$ | Optical (2) | 102.51 | 107.63 | 112.12 | × | × | × | × | × | × |
| $E_{2u}$ | Optical (2) | 189.97 | 210.45 | 224.86 | × | × | × | × | × | × |
| $E_{2u}$ | Optical (2) | 229.65 | 242.39 | 254.60 | × | × | × | × | × | × |
| $E_{2u}$ | Optical (2) | 408.81 | 439.97 | 467.26 | × | × | × | × | × | × |
| $B_{2u}$ | Optical (1) | 102.51 | 109.86 | 116.53 | × | × | × | × | × | × |
| $B_{2u}$ | Optical (1) | 259.69 | 267.97 | 275.12 | × | × | × | × | × | × |
| $B_{2u}$ | Optical (1) | 371.14 | 383.42 | 394.60 | × | × | × | × | × | × |
| $B_{2u}$ | Optical (1) | 653.73 | 683.34 | 709.78 | × | × | × | × | × | × |
| $A_{2g}$ | Optical (1) | 123.79 | 133.84 | 142.06 | × | × | × | × | × | × |
| $A_{2g}$ | Optical (1) | 186.95 | 195.94 | 203.02 | × | × | × | × | × | × |
| $A_{2g}$ | Optical (1) | 366.49 | 379.17 | 390.59 | × | × | × | × | × | × |
| $B_{1g}$ | Optical (1) | 147.13 | 154.95 | 161.78 | × | × | × | × | × | × |
| $B_{1g}$ | Optical (1) | 242.97 | 266.94 | 285.75 | × | × | × | × | × | × |
| $B_{1u}$ | Optical (1) | 174.65 | 183.25 | 190.19 | × | × | × | × | × | × |
| $B_{1u}$ | Optical (1) | 214.28 | 226.50 | 237.01 | × | × | × | × | × | × |
| $B_{2g}$ | Optical (1) | 191.71 | 196.40 | 200.47 | × | × | × | × | × | × |

The signature of vibrational spectroscopy is identified in view of the factor-group theory [61] by examining Brillouin-zone center phonon eigenvectors [87]. The irreducible representations of the vibrational modes in $Nb_5Ir_3N$ at the zone-center ($\Gamma$-center) are defined as: $\Gamma^{\text{acoustic}} = 2E_{1u} + A_{2u}$, $\Gamma^{optical}_{active} = 10E^{IR}_{1u} + 3A^{IR}_{2u} + 10E^{R}_{2g} + 2A^{R}_{1g} + 6E^{R}_{1g}$, and $\Gamma^{optical}_{silent} = 8E^{S}_{2u} + 4B^{S}_{2u} + 3A^{S}_{2g} + 2B^{S}_{1g} + 2B^{S}_{1u} + B^{S}_{2g}$, where the superscript '*IR*' and '*R*' represent infrared active and Raman active mode frequencies, respectively, while '*S*' denotes the inactive mode often known as the *silent mode*. Among all these vibrational modes, the first three modes ($2E_{1u}$ and $A_{2u}$) are acoustic with zero frequencies at the $\Gamma$-point. The remaining fifty-one optical modes consist of eighteen Raman-active and thirteen infrared-active modes, while the remaining twenty modes are optically inactive (silent). **Table 2** summarizes all the modes along with the phonon frequencies at the zone-center for $Nb_5Ir_3N$ at 0 GPa, 10 GPa, and 20 GPa. The absence of available experimental and theoretical phonon-frequency data under pressure for $Nb_5Ir_3N$ highlights the novelty of the present study. Therefore, these findings establish a useful benchmark for future theoretical investigations and experimental spectroscopic characterization regarding material's vibrational properties.

*3.1.4. Vibrational thermodynamical properties*

The thermodynamic characteristics, specifically the temperature dependence of internal energy, entropy, and heat capacity, play a critical role in estimating the thermal energy required during cycling processes. Consistent thermodynamic trends are also essential for energy-harvesting systems, where factors such as thermal regulation, heat dissipation, and temperature-dependent optical behavior significantly influence the performance of light-absorbing and photoactive materials [88].

As illustrated in **Fig. 6a**, the internal energy, free energy, and entropy of $Nb_5Ir_3N$ throughout the studied pressure remain close to zero at temperatures below 100 K. As the temperature increases, the internal energy and entropy gradually rise, whereas the free energy decreases. In view of **Fig. 6a**, the vibrational internal energy ($U_V$) manifests an approximately linear increase with temperature, indicating a continuous accumulation of thermal energy within the crystal lattice. The observed trend results from intensified atomic vibrations, leading to an increase in both the kinetic and potential energy components within the system. The near-linear dependence reflects the consistent enthalpic response of these materials, indicating their capability to store thermal energy in a stable and systematic manner under thermal excitation [83]. As seen, the internal energy reaches its maxima at 1000 K of ~3.89 eV and ~3.79 eV for 0 GPa and 20 GPa, respectively. In contrast, the negative free energy indicates that the investigated $Nb_5Ir_3N$ is thermodynamically stable [89]. The computed results also indicate that the free energy of $Nb_5Ir_3N$ displays a consistent downward trend. This observation suggests that the material maintains its structural integrity and does not undergo soft mold phase transformation or structural collapse, even at elevated temperatures. This resilience highlights its potential for applications in high-temperature environments [90]. Furthermore, the product of temperature and entropy increases more rapidly with rising temperature compared to the free energy and enthalpy. This behavior supports the idea that entropy increases with temperature as atomic thermal vibrations become stronger at higher temperatures [91]. The entropy attains maximum values of around 10.00 eV for 0 GPa and 8.67 eV for 20 GPa at 1000 K.

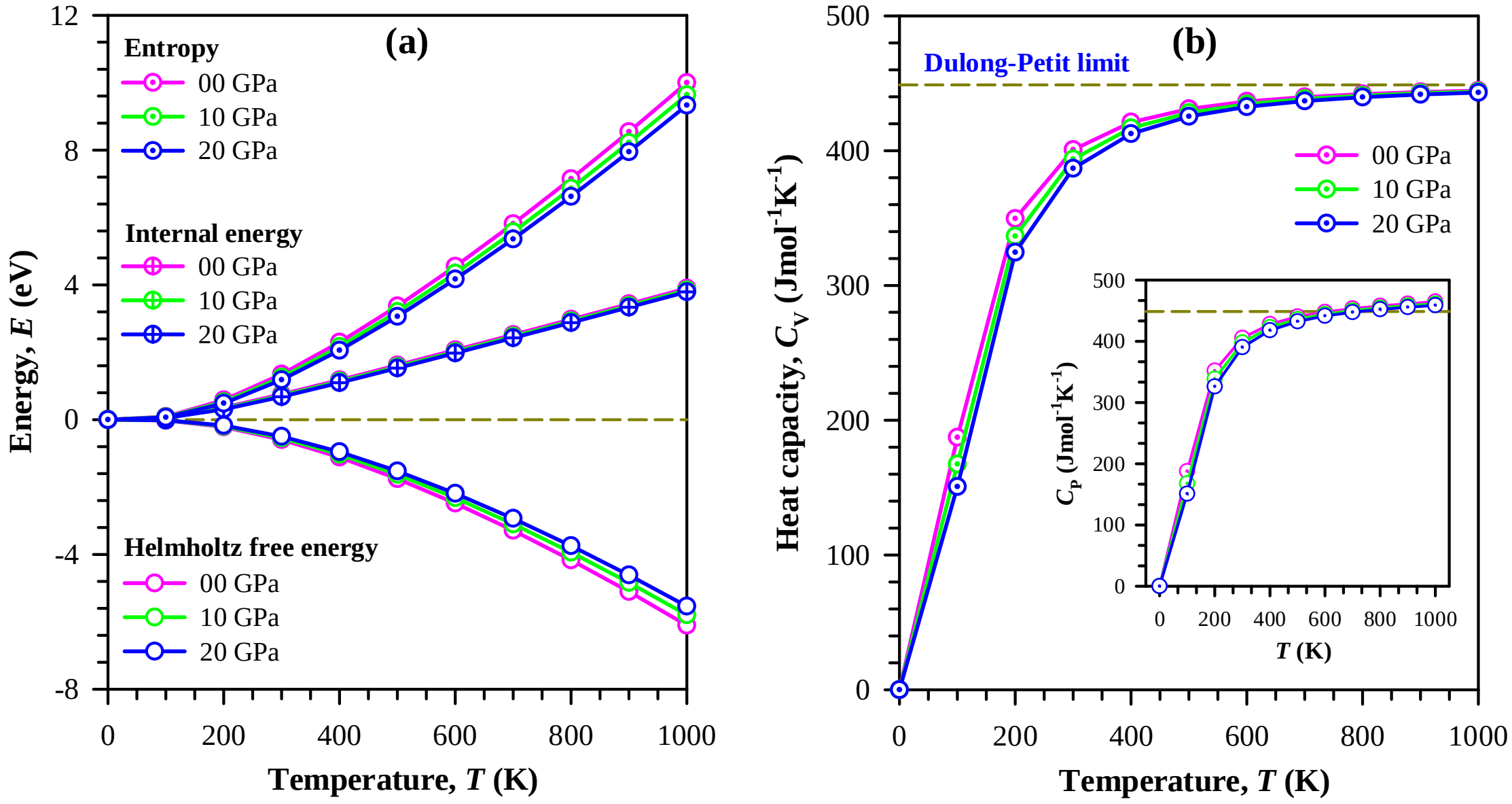

**Fig. 6**. Vibrational thermodynamic functions of $Nb_5Ir_3N$ electride as a function of temperature at pressures up to 20 GPa: (a) vibrational internal energy $U_V$, vibrational Helmholtz free energy $F_V$, vibrational entropy $S_V$, and (b) specific heat at a constant volume $C_V$.

As shown in **Fig. 6b**, the variation of the specific heat at constant volume ($C_V$) with temperature closely follows the predictions of the Debye model ($T^3$-law) reflecting that only low-frequency, long-wavelength acoustic phonon modes are active. With increasing temperature ($T \gg \Theta_D$), $C_V$ gradually approaches the classical Dulong–Petit limit [92], where it is nearly equal to $3nR$, with $R$ being the universal gas constant (8.314 J mol$^{-1}$ K$^{-1}$). This plateau arises because all phonon modes are fully excited, resulting in an even distribution of vibrational energy among all degrees of freedom. Inset of **Fig. 6b** depicts the isobaric heat capacity ($C_P$) with temperature at various pressures. As seen, temperature and pressure exert opposite effects on the heat capacity, with temperature having a more pronounced influence than pressure. The relationship between the isobaric and isochoric heat capacities is given by $(C_P - C_V) = (3\alpha)^3 BVT$, where $\alpha$ is the thermal expansion coefficient. The calculated $C_P$ values are very close to the corresponding $C_V$ values, making their difference negligible at lower temperatures [93]. At 300 K, the calculated $C_V$ ($C_P$) values are 400.65 (405.35), 393.58 (397.57), and 386.75 (390.22) Jmol$^{-1}$K$^{-1}$ for 0, 10, and 20 GPa, respectively. Therefore, both $C_P$ and $C_V$ decrease systematically with increasing pressure along 0 GPa → 10 GPa → 20 GPa.

### *3.2. Elastic constants and mechanical properties*

#### *3.2.1. Single and polycrystalline properties*

A comprehensive understanding of the mechanical properties of a material is crucial for estimating its suitability for applications. Elastic constants are fundamental parameters for evaluating the mechanical properties of solids. In addition, elastic constants provide valuable insights into the response of solids to external pressure. They are also crucial for understanding fundamental mechanical properties, including elastic anisotropy, bonding characteristics, and structural stability. The estimated elastic stiffness constants ($C_{ij}$) along with polycrystalline moduli ($B$, $G$, $Y$) for $Nb_5Ir_3N$ at various pressures are presented in **Table 3** and depicted in **Fig. 7a-b**. For a structurally stable hexagonal crystal, all five elastic constants should satisfy the following necessary and sufficient conditions [82]: $C_{44} > 0$, $(C_{11} - C_{12}) > 0$, $(C_{11} + 2C_{12})C_{33} > 2C_{13}^2$. In contrast, the stability criteria under pressure are [94]: $\tilde{c}_{44} > 0$, $\tilde{c}_{11} > |\tilde{c}_{12}|$, $\tilde{c}_{33}(\tilde{c}_{11} + \tilde{c}_{12}) > 2\tilde{c}_{13}^2$; where $\tilde{c}_{ii} = C_{ii} - P$ $(i = 1\sim4)$, $\tilde{c}_{12} = C_{12} + P$, $\tilde{c}_{13} = C_{13} + P$. Within the pressure range of 0–20 GPa, the calculated elastic constants meet the above-mentioned criteria, confirming the mechanical stability of $Nb_5Ir_3N$ throughout this pressure range.

The elastic constants $C_{11}$ and $C_{33}$ measure the resistance to linear compression along the *a*- and *c*-axes, respectively. The calculated values of $C_{11}$ and $C_{33}$ are considerably higher than those of $C_{44}$ and $C_{66}$ for all pressures which characterize the shear response of the hexagonal crystal. Furthermore, the value of $C_{66}$ is notably higher than $C_{44}$, suggests the pronounced anisotropy of shear behaviors. The elastic properties and the effect of pressure on the elastic properties of $Nb_5Ir_3N$ are reported here for the first time, constituting an important contribution to the understanding of mechanical behavior of this material. Owing to the lack of previously reported data, the calculated results offer a valuable reference for future research.

**Table 3**. Calculated single- and polycrystalline elastic constants ($C_{ij}$, $B$, $G$, $Y$ all in GPa), Poisson's ratio ($\sigma$) and Pugh's ratio ($G/B$), and Cauchy pressure [($C_{13}$-$C_{44}$) and ($C_{12}$-$C_{66}$) in GPa] for $Nb_5Ir_3N$ at various pressures.

| $P$ | $C_{11}$ | $C_{12}$ | $C_{13}$ | $C_{33}$ | $C_{44}$ | $C_{66}$ | $B$ | $G$ | $Y$ | $\sigma$ | $G/B$ | $C_{13}$-$C_{44}$ | $C_{12}$-$C_{66}$ |
|---|---|---|---|---|---|---|---|---|---|---|---|---|---|
| 00 | 441.61 | 221.20 | 171.68 | 428.07 | 75.66 | 110.21 | 270.62 | 99.29 | 265.40 | 0.337 | 0.367 | 365.95 | 111.00 |
| 10 | 492.26 | 245.88 | 197.95 | 498.30 | 88.25 | 123.19 | 307.17 | 113.23 | 302.52 | 0.336 | 0.369 | 404.01 | 122.69 |
| 20 | 551.30 | 286.98 | 230.45 | 566.36 | 99.11 | 132.16 | 351.46 | 124.66 | 334.44 | 0.341 | 0.355 | 452.19 | 154.82 |

The elastic properties of polycrystals mainly comprise the elastic moduli ($B$, $G$, and $Y$) and Poisson's ratio $\sigma$. Among them, $G$ and $B$ are calculated using VRH approximation, as given by **Eqs. 2**, while $Y$ and $\sigma$ are calculated using **Eqs. 3**. Generally, the compressibility under hydrostatic pressure is related to bulk modulus $B$; maximum $B$ indicates the minimum compressibility, $\beta = 1/B$. As seen in **Fig. 7b** and **Table 3**, the minimum $B$ (maximum $\beta$) is at 0 GPa, while the minimum $\beta$ (maximum $B$) is achieved at 20 GPa. Besides $C_{44}$ and $C_{66}$, the ability of a material to resist deformation under shear stress can also be reflected by the $G$. The greater $G$ corresponds to the stronger shear resistance of the material. Upon increasing pressure from ambient condition to 20 GPa, both $B$ and $G$ progressively increase (**Fig. 7b**). Young's modulus, which characterizes the stiffness of the material, also increases with increasing pressure.

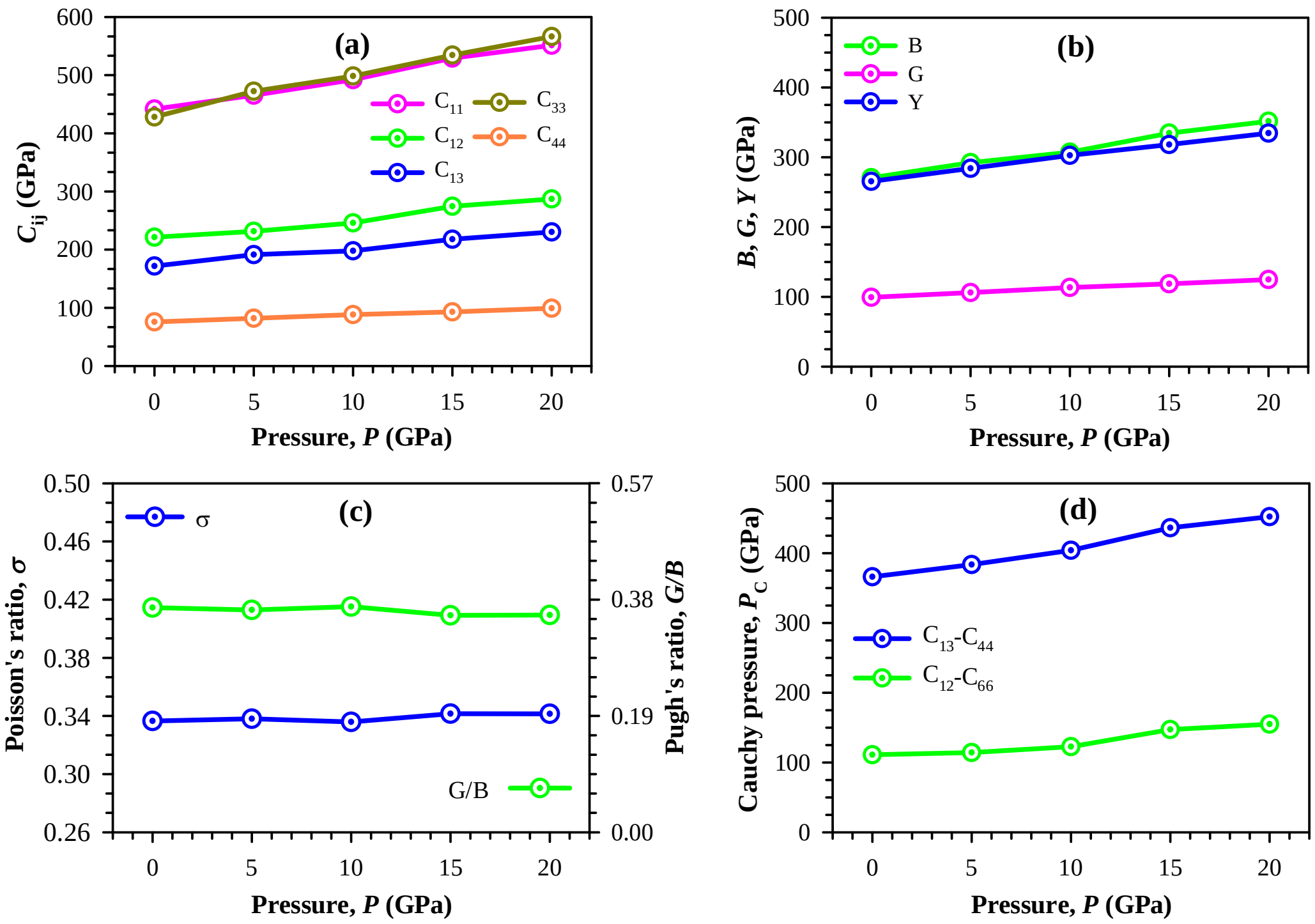

**Fig. 7**. (a) Elastic constants ($C_{ij}$), (b) moduli ($B$, $G$, $Y$), (c) Poisson's ratio ($\sigma$) and Pugh's ratio ($G/B$), and (d) Cauchy pressure ($P_C$) as a function of hydrostatic pressure for hexagonal $Nb_5Ir_3N$.

Ductility/brittleness of materials is characterized by three well-known indicators: Poisson's ratio $\sigma$, Pugh's ratio $G/B$, and Cauchy pressure $P_C$ (For hexagonal symmetry, $P_C^a = C_{13} - C_{44}$ and $P_C^b = C_{12} - C_{66}$). A material behaves as ductile if $\sigma > 0.26$, $G/B < 0.57$, and $P_C > 0$; otherwise they exhibit brittleness [95–97]. All the considered indices consistently satisfy the criteria for ductile behavior, as shown in **Fig. 7c-d** and **Table 3**, while the ductile nature becomes more pronounced with increasing pressure. In addition, Poisson's ratio can also forecast the nature of interatomic bonding forces, indicating whether they are central or non-central [98]. A Poisson's ratio within the range of 0.25–0.50 is generally associated with predominantly central interatomic forces. Based on the calculated $\sigma$, the interatomic forces in $Nb_5Ir_3N$ ($\sigma$ various from 0.336 to 0.341) are predominantly central throughout the investigated pressure range up to 20 GPa. Moreover, substantial high values (> 0.25) of $\sigma$ categorize $Nb_5Ir_3N$ as better plastic materials [99]. Cauchy pressure can also be used to predict the bonding nature of solids based on whether $P_C$ is positive or negative. The positive $P_C$ values obtained at all pressures indicate that $Nb_5Ir_3N$ exhibits metallic bonding with damage-tolerant behavior [97].

**Table 4**. Calculated machinability index ($\mu_M$), Lamé's constants ($\lambda$ and $\mu$), Kleinmann parameter ($\xi$), Vicker's hardnesses ($H_V$ in GPa), fracture toughness ($K_{IC}$ in MPa m$^{1/2}$), critical energy release rate ($G_{IC}$ in Jm$^{-2}$), linear compressibility coefficients ($f$), universal anisotropy index ($A^U$ in GPa), and anisotropy in elastic moduli ($A_Y$, $A_\beta$, $A_G$, and $A_\sigma$) for $Nb_5Ir_3N$ at various pressures.

| $P$ | $\mu_M$ | $\lambda$ | $\mu$ | $\Xi$ | $H_V^{Miao}$ | $H_V^{M-O}$ | $K_{IC}$ | $G_{IC}$ | $f$ | $A^U$ | $A_Y$ | $A_\beta$ | $A_G$ | $A_\sigma$ |
|---|---|---|---|---|---|---|---|---|---|---|---|---|---|---|
| 00 | 3.577 | 204.43 | 99.29 | 0.668 | 10.82 | 14.75 | 2.58 | 22.22 | 0.239 | 0.294 | 1.52 | 1.25 | 1.72 | 2.17 |
| 10 | 3.481 | 231.68 | 113.23 | 0.666 | 12.39 | 16.80 | 2.92 | 24.95 | 0.146 | 0.262 | 1.51 | 1.14 | 1.66 | 2.05 |
| 20 | 3.546 | 268.35 | 124.66 | 0.687 | 13.18 | 18.68 | 3.26 | 28.03 | 0.123 | 0.253 | 1.52 | 1.12 | 1.63 | 1.93 |

The machinability index ($\mu_M$) is an important mechanical parameter for assessing the engineering applicability of solids. Machinability refers to the ease and efficiency with which a solid can be processed or shaped using cutting tools and it influences various machining parameters, including cutting force, cutting energy, drilling rate, feed rate, tool wear, machining time, and depth of cut [**100**,**101**]. Moreover, it influences the flexibility and dry lubricating properties of crystals. A higher machinability index indicates greater ease of machining. The machinability index, $\mu_M$ of a material is defined as, $\mu_M = B/C_{44}$. The calculated $\mu_M$ are listed in **Table 4**, showing a non-monotonic pressure dependence throughout the investigated range; showing a maximum machinability at 0 GPa (~ 3.577). The compound under investigation is highly machinable and has excellent dry lubricity.

Kleinman parameter ($\xi$), a dimensionless strain parameter, is a measurement that estimates the stability of a compound against bond bending and stretching. It can be calculated as: $\xi = (C_{11} + 8C_{12})/(7C_{11} + 2C_{12})$ [**100**]. Generally, $\xi$ lies within zero and one as: $0 \leq \xi \leq 1$. Typically, a low value of $\xi$ indicates strong resistance to bond bending and vice-versa, *i.e.*, the bond bending (stretching) will be maximum when $\xi = 1$ ($\xi = 0$). As in **Table 4**, the estimated $\xi$ reaches its maximum at 20 GPa (~ 0.687), suggesting it is influenced by bond stretching or shrinkage at this pressure [**100**].

Lamé coefficients ($\lambda$ and $\mu$) help to understand how solids respond to stress and strain. In general, $\lambda$ measures the compressibility and $\mu$ quantifies the shear stiffness of material [**102**]. They are computed as: $\lambda = Y\sigma/(1+\sigma)(1-2\sigma)$ and $\mu = Y/2(1+\sigma)$. It is evident that both are directly proportional to $Y$ in which $\mu$ is nothing more than $G$ (cf. **Table 3** and **Table 4**); that is, $\mu = G$. Moreover, for isotropic solid, the conditions are: $\lambda = C_{12}$ and $\mu = (C_{11}\text{-}C_{12})/2$ [**103**]. The material exhibits anisotropic behavior, as the second coefficient fails to meet the above-mentioned criterion. As shown in **Fig. 8a**, the pronounced increase in $\lambda$ and $\mu$ with pressure suggests that $Nb_5Ir_3N$ becomes increasingly tough at elevated pressures.

*3.2.2. Hardness and fracture toughness*

Theoretical Vickers hardness ($H_V$) is an important mechanical property that quantifies a material's resistance to permanent (plastic) deformation and is commonly estimated from elastic properties using various empirical models. In this work, two frequently-used methods, proposed by Miao [**104**] and Mazhnik-Oganov (M-O) [**105**], are used to estimate Vicker's hardness of $Nb_5Ir_3N$ as:

$$\left.\begin{aligned} H_V^{\text{Miao}} &= \frac{(1-2\sigma)Y}{6(1+\sigma)}, \\ H_V^{M-O} &= \frac{0.096(1-8.5\sigma+19.5\sigma^2)Y}{1-7.5\sigma+12.2\sigma^2+19.6\sigma^3}. \end{aligned}\right\} \tag{13}$$

Fracture toughness ($K_{IC}$) quantifies a material's resistance to crack propagation and is one of the most important mechanical properties for assessing structural reliability. It can be calculated by Niu's empirical model as [**106**]:

$$K_{IC} = V_0^{\frac{1}{6}} G \left(\frac{B}{G}\right)^{\frac{1}{2}}, \tag{14}$$

where $V_0$ is the atomic volume in m$^3$, and $G$ and $B$ are expressed in MPa.

The critical energy release rate ($G_{IC}$), analogous to fracture toughness ($K_{IC}$), represents the energy needed to move a crack. It is calculated as a function of fracture toughness ($K_{IC}$) as [107]:

$$G_{IC} = K_{IC}^2 \left(\frac{1-\sigma^2}{Y}\right). \tag{15}$$

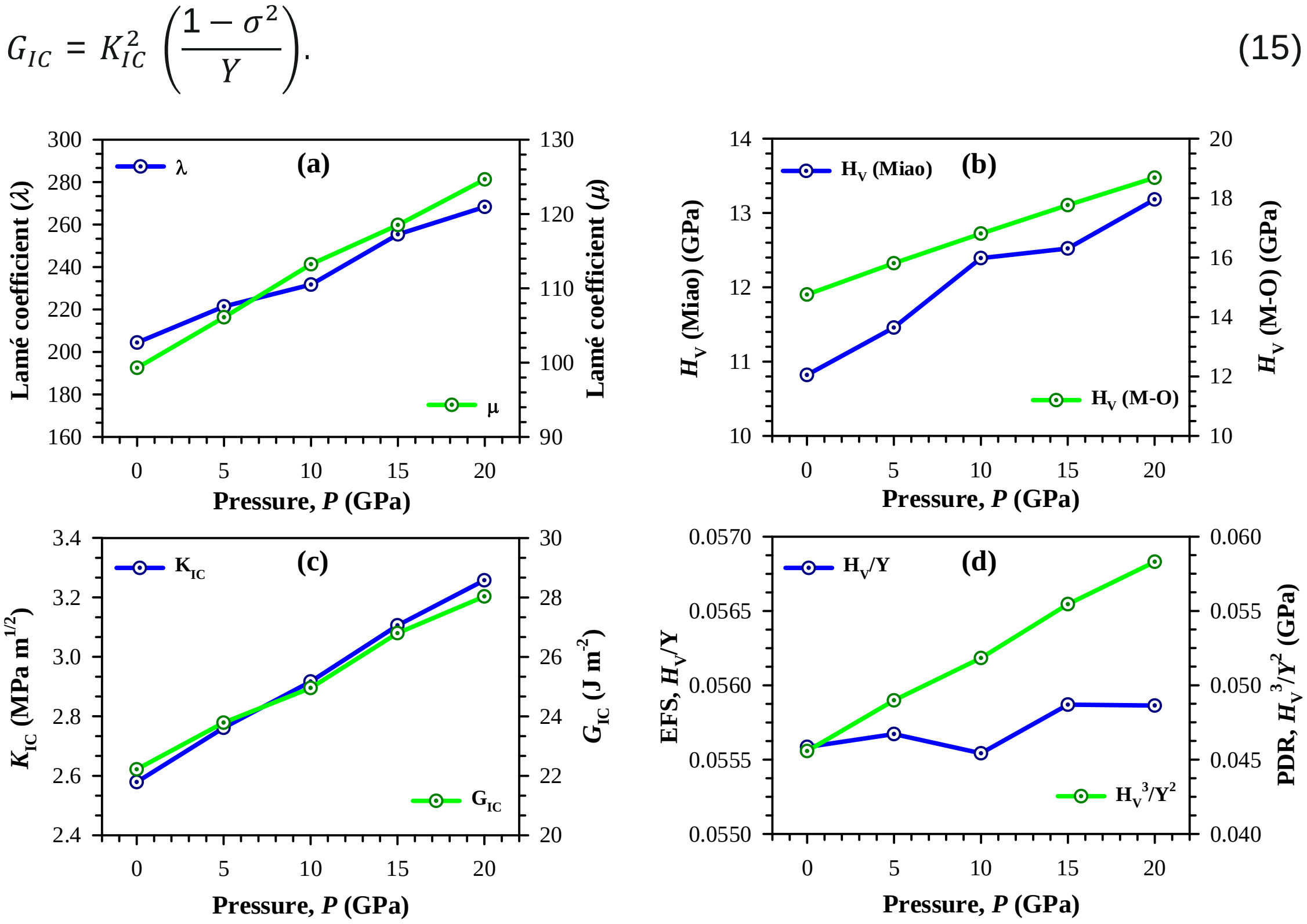

**Fig. 8**. (a) Lamé coefficients ($\lambda$ and $\mu$), (b) Vickers harnesses ($H_V$), (c) Fracture toughness ($K_{IC}$) and critical energy release rate ($G_{IC}$), and (d) elastic failure strain (EFS) and plastic deformation resistance (PDR) as a function of pressure for $Nb_5Ir_3N$.

The calculated Vickers hardness increases from 10.8 to 13.2 GPa (Miao model) and 14.8 to 18.7 GPa (Mazhnik–Oganov model) as pressure increases from 0 to 20 GPa (cf. **Table 4** and **Fig. 8b**), indicating pressure-induced strengthening of atomic bonding. Since the hardness exceeds the 10 GPa threshold for hard materials [108], $Nb_5Ir_3N$ can be classified as a hard material with enhanced mechanical robustness under compression, making it a promising candidate for high-pressure and wear-resistant applications. Generally, the higher hardness exhibits the better the wear resistance [109]. To estimate the wear resistance of materials, Leyland and Matthews [110] proposed two factors: elastic failure strain (EFS), $H_V/Y$ and plastic deformation resistance (PDR), $H_V^3/Y^2$. As seen in **Fig. 8d**, the calculated EFS and PDR increase with increasing pressure, reaching their maximum values at 20 GPa, which indicates the strongest wear resistance at this pressure.

According to the calculated fracture toughness $K_{IC}$ (**Table 4** and **Fig. 8c**), the lowest $K_{IC}$ is achieved at ambient pressure (~ 2.58 MPa $m^{1/2}$) and maximum at 20 GPa (~ 3.26 MPa $m^{1/2}$), indicating that applying pressure enhance the intrinsic crack propagation resistance of the material. However, in hard materials, crack initiation during machining generally occurs under localized sharp-contact conditions generated by abrasive grains or cutting edges, and is governed not only by fracture toughness but also strongly by hardness [**111**]. According to Lawn–Evans model, the critical load (CL) to initiate crack is expressed as [**112**]: $P^{CL} \propto ({K_{IC}}^4/{H_V}^3)$. Therefore, the relative critical crack initiation (RCCI) load, defined as $(P^{CL}/P_0^{CL})$, is used to evaluate the machining-induced crack initiation proclivity within 0-20 GPa for $Nb_5Ir_3N$. In view of computed results, RCCI loads are all higher than that of 0 GPa (1.00) due to the substantial increment in $K_{IC}$ and hardness (used $H_V^{M-O}$ value). The RCCI loads increase from 1.11 to 1.25 at 10 and 20 GPa, respectively, indicating that a higher load is required to initiate cracks under identical localized sharp-contact conditions. This behavior suggests enhanced resistance to brittle cracking and subsurface damage during machining [**111**]. Similar to the fracture toughness ($K_{IC}$), the critical energy release rate ($G_{IC}$) exhibits a nearly identical increasing trend with pressure (cf. **Fig. 8c**), reaching a maximum value of 28.03 $Jm^{-2}$ at 20 GPa, which indicates enhanced resistance to crack propagation.

*3.2.3. Elastic anisotropy*

Elastic anisotropy is a fundamental aspect influencing the mechanical durability and the predicted microhardness of materials, as it governs the directional dependence of their mechanical response under applied stress. A pronounced elastic anisotropy may lead to the initiation and propagation of microcracks, adversely affecting the mechanical integrity of the material [**113**]. To characterize the pressure-induced variations in elastic anisotropy, the universal elastic anisotropy index ($A^U$) is adopted as a comprehensive descriptor for the investigated compound, which is evaluated as [**114**]:

$$A^U = \left(\frac{5G_V}{G_R} + \frac{B_V}{B_R} - 6\right) \geq 0. \quad (16)$$

For a perfectly isotropic system, $A^U$ have to be zero, and any value greater than zero measures the strength of anisotropy. That is, the smaller the value of $A^U$, the closer the material is to isotropic behavior. The calculated values of $A^U$ are 0.294, 0.262, and 0.253 at 0 GPa, 10 GPa, and 20 GPa, respectively. The universal anisotropy index decreases with pressure, indicating a gradual reduction in elastic anisotropy. This pressure-induced behavior suggests that $Nb_5Ir_3N$ becomes progressively more isotropic upon compression, exhibiting a more consistent elastic response across various crystallographic orientations.

For a hexagonal system, the ratio of linear compressibility coefficients along *c*-to-*a* axis serves as an important measure to study the anisotropic nature of crystal, which is defined as [**115**],

$$f = \frac{k_c}{k_a} = \frac{C_{11} + C_{12} - 2C_{13}}{C_{33} - C_{13}}. \quad (17)$$

Isotropic compressibility is achieved when $f = 1$, while the deviation from the unity measures the degree of anisotropy [116]. The calculated values of $f$ are 0.239, 0.146, and 0.123 for 0 GPa, 10 GPa, and 20 GPa, respectively, exhibits the enhancement of isotropic nature while going to higher pressures from ambient.

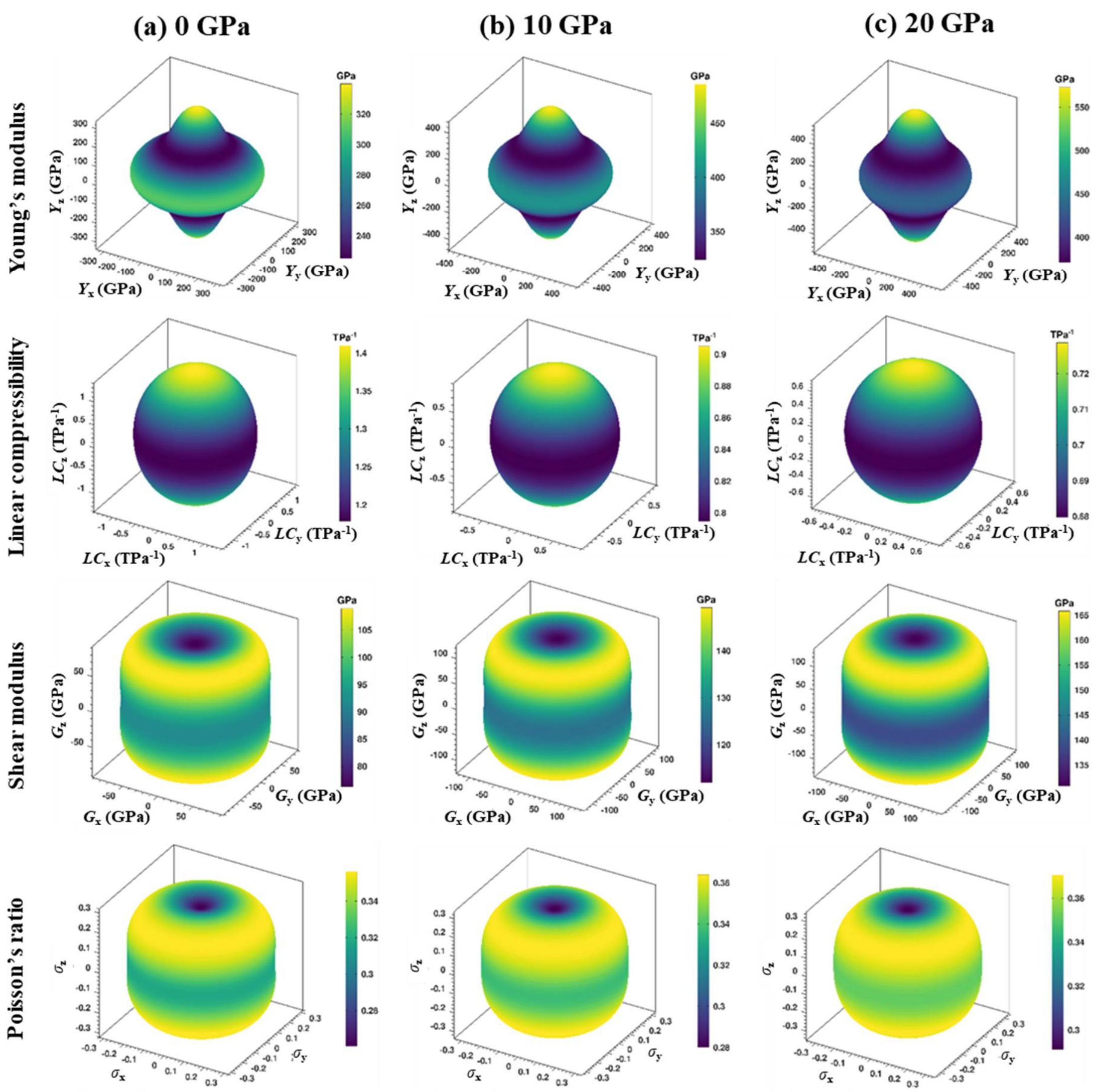

**Fig. 9**. Directional dependences of Young's modulus, linear compressibility, Shear modulus, and Poisson's ratio of $Nb_5Ir_3N$ at (a) 0 GPa, (b) 10 GP, and (c) 20 GPa.

To get a more comprehensive insight into the anisotropic features of $Nb_5Ir_3N$, we have utilized the VELAS code [117] to visualize three-dimensional (3D) plots by means of elastic parameters, namely Young's modulus, linear compressibility, shear modulus, and Poisson's ratio. **Figure 9** represents the directional dependences of the elastic moduli of $Nb_5Ir_3N$ at 0 GPa, 10 GP, and 20 GPa. When the plot takes on a spherical shape in three dimensions, it unmistakably demonstrates a state of complete isotropy. Any deviations from this spherical form clearly indicate the varying degrees of anisotropy present in different orientations within the three-dimensional space. The calculated Young's modulus, linear compressibility, shear modulus, and Poisson's ratio of materials exhibit significant anisotropy in all the planes. As shown in **Table 4**, the strength of elastic anisotropy decreases progressively with increasing pressure, in agreement with the behavior of the other anisotropy indices.

### *3.3. Thermo-mechanical properties*

#### *3.3.1. Debye and melting temperatures*

Acoustic sound velocities are the significant parameters which can reflect the thermal properties of materials in solids. These velocities, namely, longitudinal ($v_l$) and transverse sound velocity ($v_t$), may be estimated by means of Navier's equation [**118**]:

$$v_l = \left(\frac{3B + 4G}{3\rho}\right)^{\frac{1}{2}};\ v_t = \left(\frac{G}{\rho}\right)^{\frac{1}{2}};\ v_m = \left[\frac{1}{3}\left(\frac{2}{v_t^3} + \frac{1}{v_l^3}\right)\right]^{-\frac{1}{3}}, \tag{18}$$

where $v_m$ is the average sound velocity and $\rho$ is the density of the crystal. By utilizing **Eqn. 19**, we may obtain another fundamental physical parameter, the Debye temperature $\Theta_D$, which is closely related to many physical properties including specific heat and melting temperature. Debye temperature is expressed using the Anderson model [**119**]:

$$\Theta_D = \frac{h}{k_B}\left[\frac{3n}{4\pi}\left(\frac{N_A\rho}{M}\right)\right]^{\frac{1}{3}} v_m, \tag{19}$$

where $h$ and $k_B$ are the Planck's and Boltzmann's constants, respectively, $N_A$ is the Avogadro's number, $M$ is the molecular weight, and $n$ is the number of atoms contained in one molecular formula.

The melting temperature ($T_m$) is an important parameter for evaluating the suitability of solids for high-temperature applications. Knowledge of $T_m$ is essential for the rational design and performance optimization of high-temperature materials. Following the relation proposed by Fine *et al.*, $T_m$ for hexagonal solid is calculated as [**120**]: $T_m = 354 + 1.5(2C_{11} + C_{33})$.

As seen in **Table 5** and **Fig. 10a**, the sound velocities and Debye temperature all intensified with increasing pressures; therefore, the increment of pressure indicates the hardening with pressure of $Nb_5Ir_3N$ as evident by Vickers hardness $H_V$. The calculated Debye temperature ranges from ~ 309 K (at 0 GPa) to ~ 343 K (at 20 GPa).

The calculated melting temperature escalates from 3163 to 3811 K (**Fig. 10b**) as the pressure increases from 0 to 20 GPa, indicating that $Nb_5Ir_3N$ possesses an increasingly stiff lattice and enhanced thermal stability under compression. Since $T_m$ is forecasted from the elastic constants, particularly from $C_{11}$ and $C_{33}$, its increment with pressure is consistent with pressure-induced enhancement of the axial elastic stiffness. Increasing pressure promotes atomic densification, thereby intensifying interatomic interactions and driving a systematic shortening of equilibrium bond lengths. Consequently, more thermal energy is required to disrupt the crystal structure, leading to a higher melting temperature.

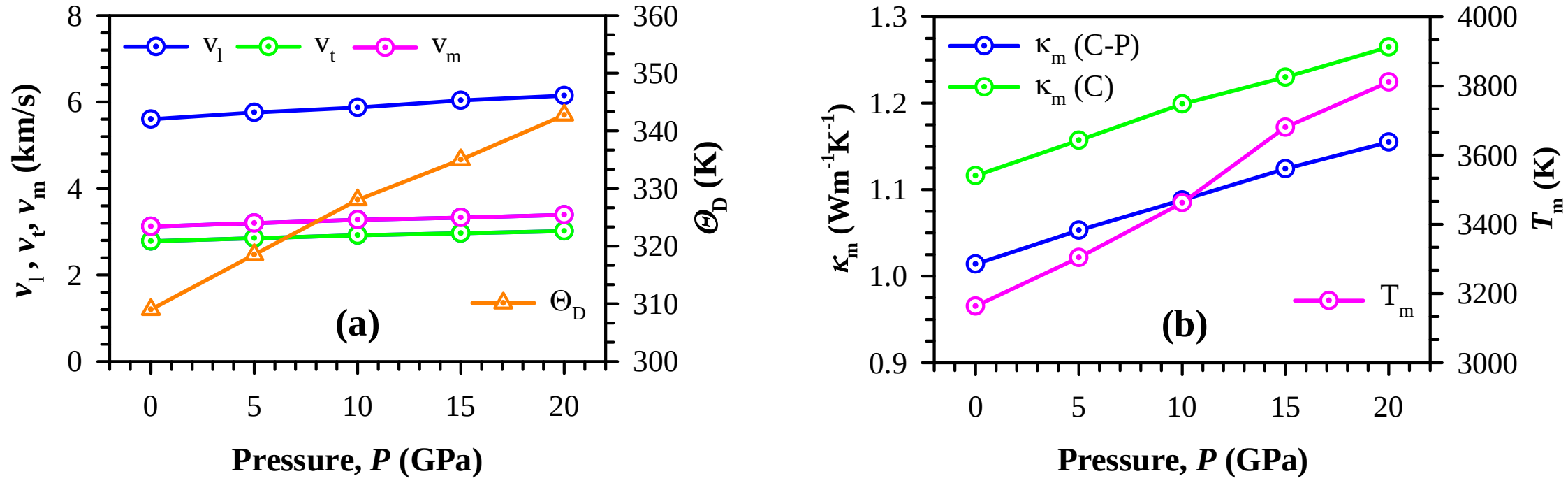

**Fig. 10**. Calculated (a) Longitudinal, transverse and average ($v_l$, $v_t$, and $v_m$ in km/s) sound velocities and Debye temperatures ($\Theta_D$ in K), and (b) minimum thermal conductivities ($\kappa_m$ in $Wm^{-1}K^{-1}$) and melting temperatures ($T_m$ in K) as a function of pressure for $Nb_5Ir_3N$.

**Table 5**. Calculated longitudinal, transverse and average ($v_l$, $v_t$, and $v_m$ in km/s) sound velocities, Debye temperatures ($\Theta_D$ in K), melting temperatures ($T_m$ in K), lattice thermal conductivities ($\kappa_l$ in $Wm^{-1}K^{-1}$), minimum thermal conductivities ($\kappa_m$ in $Wm^{-1}K^{-1}$) for $Nb_5Ir_3N$ electride at various pressures ($P$ in GPa).

| $P$ | $v_l$ | $v_t$ | $v_m$ | $\Theta_D$ | $T_m$ | $\kappa_L$ | $\kappa_m^{C-P}$ | $\kappa_m^{C}$ |
|---|---|---|---|---|---|---|---|---|
| 0 | 5.602 | 2.781 | 3.121 | 309.00 | 3162.9 | 3.050 | 1.014 | 1.116 |
| | | | | 327.0[a] | | | | |
| 10 | 5.870 | 2.918 | 3.275 | 328.04 | 3462.3 | 4.235 | 1.088 | 1.199 |
| 20 | 6.147 | 3.016 | 3.388 | 342.73 | 3811.2 | 4.945 | 1.155 | 1.265 |

[a] [18]

### *3.3.2. Minimum and lattice thermal conductivities*

Thermal conductivity ($\kappa$) is an essential parameter for assessing the heat-conducting ability of solids. To evaluate the suitability of a material for thermal barrier coating applications, its thermal conductivity needs to be investigated. Particularly in high-temperature applications, the minimum thermal conductivity ($\kappa_m$) is particularly important as a limiting parameter. To evaluate $\kappa_m$ of a material two empirical model are used: Cahill and Pohl [121] model and Clarke model [122]:

$$\left.\begin{aligned} \kappa_m^{C-P} &= \frac{1}{2.48}\, k_B n_v^{\frac{2}{3}} (v_l + 2v_t), \\ \kappa_m^{C} &= k_B v_m \left(\frac{M}{n\rho N_A}\right)^{-\frac{2}{3}}, \end{aligned}\right\} \tag{20}$$

where $n_v$, $n$, and $k_B$ denotes the number of atoms per unit volume, the number of atoms in the unit cell, and Boltzmann's constant, respectively.

The lattice thermal conductivity ($\kappa_L$) quantifies the heat carried by lattice vibrations in response to a temperature gradient within a solid. Its temperature dependence can be evaluated using the empirical model proposed by Slack as [123]:

$$\kappa_L = A\frac{M_{av}\Theta_D^3\delta}{\gamma_a{}^2 n^{\frac{2}{3}} T}, \tag{21}$$

where $M_{av}$ is the average atomic mass in kg/mol, $\delta$ is the cubic root of the average atomic volume in meter, and $\gamma_a$ refers to the acoustic Grüneisen parameter which is a function of Poisson's ratio ($\sigma$): $\gamma_a = 1.5(1+\sigma)/(2-3\sigma)$, and $T$ is the absolute temperature in K, and the factor $A$ ($\gamma_a$) is a constant which can be calculated according to Julian as [124]: $A(\gamma_a) = 4.85628 \times 10^7 / 2(1 - 0.514\gamma_a^{-1} + 0.228\gamma_a^{-2})$.

The calculated isotropic minimum thermal conductivities ($\kappa_m$) for two different methods and lattice thermal conductivities ($\kappa_L$) are listed in **Table 5** and **Fig. 11b**. It is observed that $\kappa_m$ values escalate along 0 GPa → 20 GPa in both models, forecasting the highest hardness exhibited at 20 GPa in $Nb_5Ir_3N$. At the ground state, the lattice contribution is pivotal to the thermal conductivity, and a decrease in thermal conductivity is generally accompanied by a lower Debye temperature. The observed trend is in good agreement with the Callaway–Debye theory [125]. Moreover, the low minimum thermal conductivity predicted at ambient pressure demonstrate the potential of these materials for use as thermal barrier coatings [86].

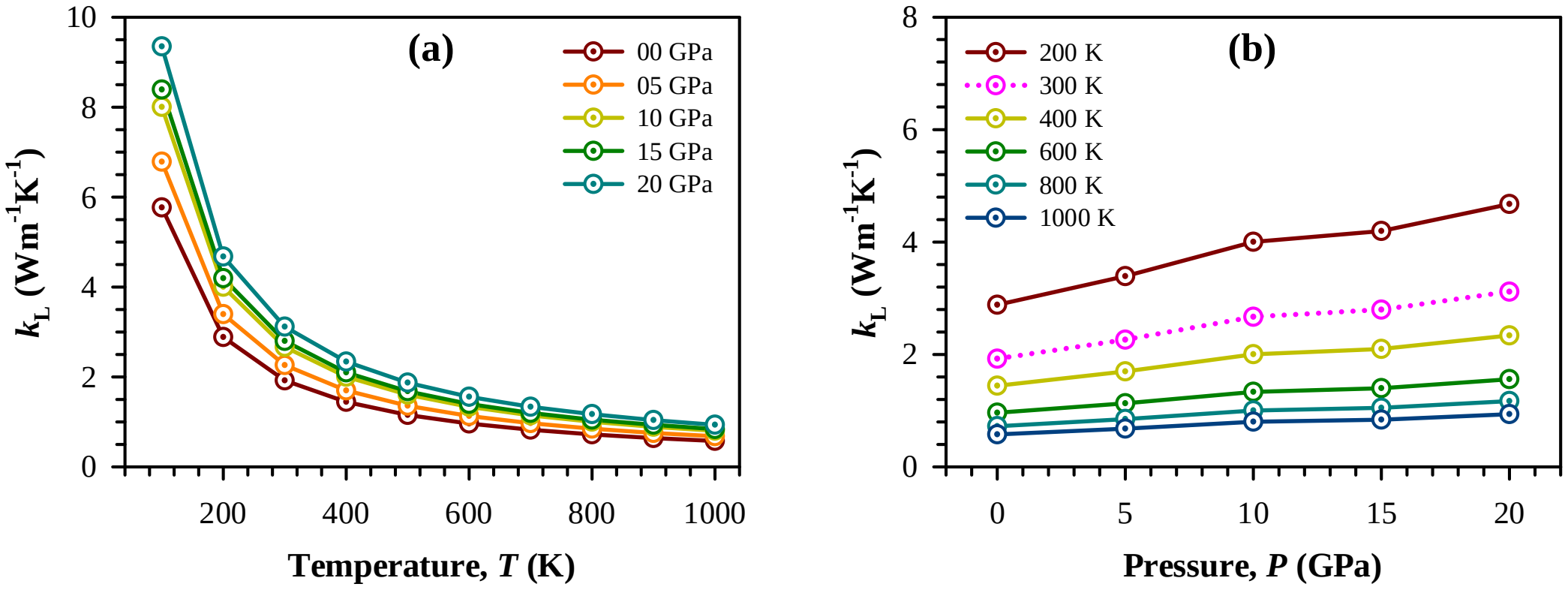

**Fig. 11**. (a) Variation of lattice thermal conductivity with temperature at various pressure and (b) Variation of lattice thermal conductivity with pressure at various temperature for $Nb_5Ir_3N$.

As conventional DFT calculations are based on the ground-state condition of 0 K, they do not fully represent the behavior of solids under realistic environmental conditions. Therefore, the influence of temperature on the properties of solids should be taken into consideration. Consequently, the temperature-dependent (from 100 K to 1000 K) lattice thermal conductivities at various pressures are calculated for $Nb_5Ir_3N$. The temperature and pressure dependence of $Nb_5Ir_3N$ on the $\kappa_L$ is shown in **Fig. 11a** and **Fig. 11b**, respectively. In view of **Fig. 11a**, a gradual decrease in lattice thermal conductivity is observed as temperature increases at each pressure, whereas **Fig. 11b** displays a monotonically increasing trend of $\kappa_L$ with pressure at a particular temperature. It is noteworthy that $\kappa_L$ tends to approach a saturation value at elevated temperatures. In contrast, at room-temperature (RT = 300 K), $\kappa_L$ is calculated as 1.922 and 3.115 Wm$^{-1}$K$^{-1}$ for 0 and 20 GPa, respectively.

### *3.4. Electronic properties*

#### *3.4.1 Band structure and density of states*

Band structure is used to study the electronic properties of materials in condensed matter physics and materials science. They describe the variation of electron energy levels as a function of crystal momentum (*k*-vector) within the Brillouin zone. By solving the electronic structure equations, usually through Density Functional Theory (DFT), these calculations provide information about the allowed energy bands, band gaps, Fermi level, and density of electronic states. Band structure calculations also provide insights into the dispersion characteristics of a material by identifying the dominant orbital character of the bands near the Fermi level ($E_F$), which is conventionally set to zero. Band structure calculations are essential for predicting opto-electronic, magnetic, and superconducting properties, making them a powerful tool for designing and understanding materials.

In this study, the band structure of $Nb_5Ir_3N$ is calculated within the Quantum Espresso code [**126**] at ambient and high pressures considering spin–orbit coupling (SOC) effect. It is noteworthy that the inclusion or exclusion of SOC is an important aspect for accurately describing the electronic nature of materials, especially materials containing heavy elements with large atomic numbers. SOC includes relativistic effects arising from the heavy Ir and Pd atoms. SOC can cause band splitting, modify band gaps, change the position of band extrema, and strongly influence the Fermi surface. Band structure of $Nb_5Ir_3N$ are also conducted with non-SOC to compare with the SOC effect although non-SOC neglect the interaction between an electronic spin and its orbital motion [**127**].

In the calculated band structures (seen in **Fig. 12**), the inclusion of SOC (blue dashed curves) produces noticeable band splitting and energy shifts compared with the non-SOC calculations (green solid curves) at both ambient and high pressure. These observed deviations originate from the relativistic interaction between the electron spin and its orbital motion, which is particularly pronounced for compounds containing heavy elements such as Nb and Ir because of their large atomic numbers and strong relativistic effects. Under compression, the reduced interatomic distances enhance orbital hybridization and broaden the electronic bandwidth, thereby modifying the magnitude of SOC-induced splitting near $E_F$. The SOC interaction can lift band degeneracies along high-symmetry directions, alter the curvature of electronic bands, and redistribute the density of states around $E_F$, leading to pressure-dependent changes in electronic transport and possible topological features. Similar pressure-enhanced SOC-driven band evolution has been reported in heavy-element intermetallic and transition-metal compounds, where applied pressure tunes the balance between electronic bandwidth and relativistic interactions [**128**].

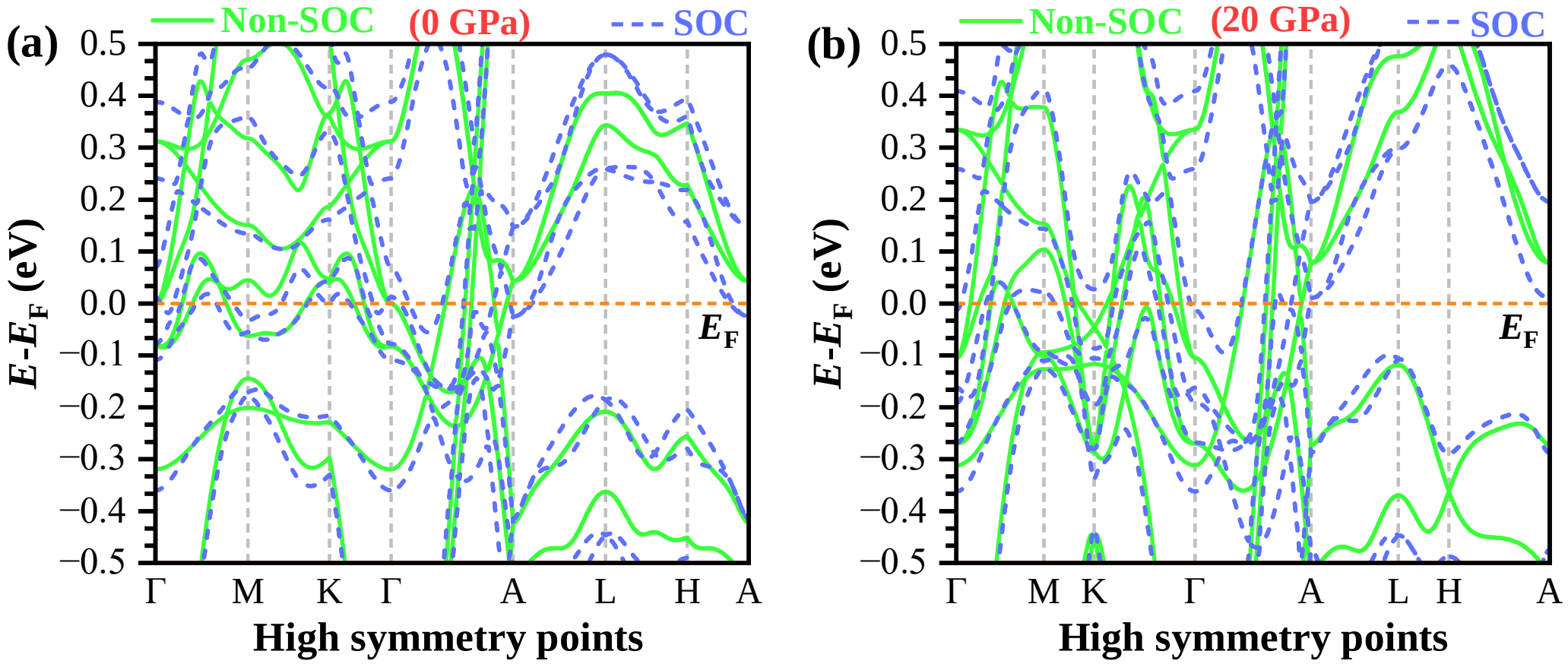

**Fig. 12**. Electronic band structures of $Nb_5Ir_3N$ at (a) 0 GPa and (b) 20 GPa hydrostatic pressure.

The electronic band structure of $Nb_5Ir_3N$ at ambient pressure (0 GPa), shown in **Fig. 12**, is in good agreement with the first-principles results reported in the reference studied by Yang *et al.* [18]. The calculated bands in the reference paper demonstrate that $Nb_5Ir_3N$ is a three-dimensional metallic system, with multiple bands crossing the Fermi level ($E_F$). The present band structure at 0 GPa also exhibits several dispersive bands crossing $E_F$ along the high-symmetry directions *Γ-M-K-Γ-A-L-H-A*, confirming the absence of an electronic band gap and the metallic nature of this compound. In both the reported and present calculations, the states near $E_F$ are primarily governed by the hybridization of Nb-4*d* and Ir-5*d* orbitals, with a smaller contribution from N-2*p* states (further confirmed by density of states calculation). The strong dispersion of these bands indicates considerable orbital overlap and relatively low effective carrier masses [129], which is consistent with the good electronic conductivity expected for $Nb_5Ir_3N$. Furthermore, $Nb_5Ir_3N$ as a nitrogen-filled electride, where interstitial electrons contribute to the electronic structure and coexist with conventional metallic bands.

A direct comparison between the non-SOC (green solid lines) and SOC (blue dashed lines) calculations at 0 GPa reveals that SOC mainly modifies the fine electronic structure rather than changing the overall metallic character. This behavior is consistent with the earlier study [18], where SOC calculations were found to introduce noticeable band splitting because of the strong relativistic interaction associated with heavy Ir atoms [130]. At ambient pressure, the largest SOC-induced changes appear near band crossings and high-symmetry regions where bands are closely spaced. These splitting remove some accidental degeneracies and slightly shift the bands relative to $E_F$; however, no SOC-induced gap opening occurs, indicating that $Nb_5Ir_3N$ remains a metal. The SOC effect is therefore a tuning mechanism for the Fermi surface rather than a driving force for a metal–semiconductor or metal–insulator transition. Such behavior is typical for transition-metal compounds containing 5*d* elements [131], where SOC modifies electronic states near $E_F$ while preserving the fundamental band topology.

Compared with the ambient-pressure case, the band structure at 20 GPa exhibits enhanced band dispersion, especially along the *Γ-M-K-Γ* and *A-L-H-A* directions. The increased slope of several bands crossing $E_F$ indicates modified carrier dynamics and a reduction in the effective mass of some charge carriers. Therefore, pressure acts to increase electronic delocalization and strengthens the metallic character of $Nb_5Ir_3N$. The pressure-induced redistribution of electronic states near $E_F$ changes the Fermi surface geometry and may influence the electron–phonon coupling responsible for superconductivity. However, the absence of a band gap at both 0 and 20 GPa indicates that pressure does not induce a semiconductor transition; instead, it continuously tunes the metallic electronic structure. The preserved metallicity together with enhanced band dispersion suggests that $Nb_5Ir_3N$ can maintain efficient charge transport under compression.

The calculated density of states (DOS) of $Nb_5Ir_3N$ at ambient pressure (0 GPa), shown in the **Fig. 13**, is consistent with the electronic structure reported by Yang *et al.* [**18**]. The total density of states (TDOS) demonstrates a finite value at the $E_F$, confirming the metallic nature of $Nb_5Ir_3N$. $E_F$ resides close to a valley in the TDOS, reflecting robust electronic structure avoiding strong electronic instabilities that could arise from an excessive enhancement of DOS. Such environment is favorable for conventional superconductivity as it provides adequate electronic states for effective electron–phonon coupling while avoiding competing electronic instabilities.

The PDOS analysis of $Nb_5Ir_3N$ reveals the electronic states around $E_F$ are mainly governed by Nb-4*d* and Ir-5*d* orbital hybridizations, whereas N-2*p* orbitals contribute negligibly in this energy region. The significant spectral weight of Nb-*d* states spanning a broad region across $E_F$ highlights the dominant influence of Nb atoms on the conduction behavior. Conversely, the Ir-*d* orbitals contribute strongly in the deeper valence region (approximately -6 to -2 eV) and also provide noticeable states near $E_F$. Such strong hybridization of Nb-*d*/Ir-*d* reflects the metallic bonding characteristics of $Nb_5Ir_3N$. The N-2*p* states primarily contribute to the lower valence bands through hybridization with transition-metal *d* orbitals, revealing the role of nitrogen in tuning the chemical bonding environment, whereas the states responsible for conduction remain mainly derived from transition-metal orbitals. The observed orbital characteristics are consistent with the nitrogen-filled electride nature of $Nb_5Ir_3N$, demonstrating that metallic conductivity arises from the interplay between interstitial electrons and *d*-state contributions of transition-metal.

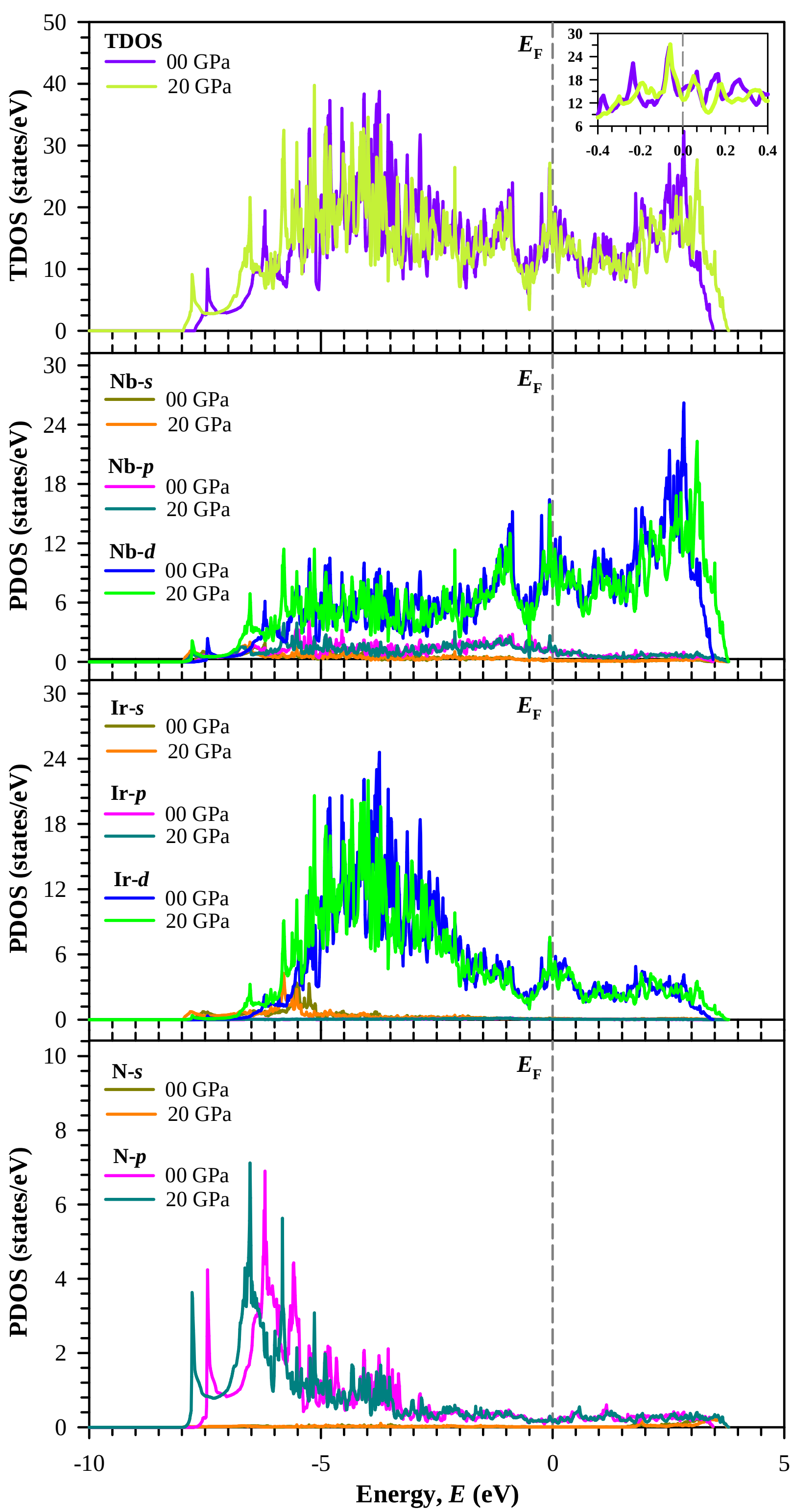

**Fig. 13**. Electronic density of states of $Nb_5Ir_3N$ at various pressures.

The calculated TDOS at $E_F$ is 7.80, 6.78, and 6.40 states $eV^{-1}$ $f.u.^{-1}$ for 0, 10, and 20 GPa, respectively. That is, TDOS of $Nb_5Ir_3N$ diminishes at higher pressures. Among the different orbital contributions, the transition-metal *d* states exhibit the strongest response to applied pressure. The redistribution and increased intensity of Nb-*d* states near $E_F$ manifest enhanced Nb-centered orbital interactions and hybridization induced by lattice compression. Likewise, the Ir-*d* states exhibit pressure-induced spectral redistribution, indicating increased overlap among the extended 5*d* orbitals. Due to the relatively delocalized character of Ir orbitals, their electronic contributions are particularly sensitive to compression. The enhanced Nb-4*d*/Ir-5*d* interaction modifies the electronic states responsible for conduction and alter the Fermi surface characteristics observed in the band structure.

### *3.4.2. Mulliken population and bonding features*

Analysis of Mulliken atomic population is useful for investigating the bonding characteristics, charge-transfer mechanism, and effective valence charge (EVC) of materials. The EVC is calculated as the difference between the formal ionic charge and the atomic charge obtained from Mulliken population analysis (MPA). An EVC value of zero corresponds to an ideal ionic bond, whereas any deviation from zero indicates the presence of covalent bonding. Moreover, a larger absolute value of EVC reflects a greater degree of covalency in the chemical bond [**132**]. All bands spilling parameter for Mulliken analysis are 0.12%, 0.12%, 0.13% for 0 GPa, 10 GPa, and 20 GPa, respectively. The obtained spilling parameters for MPA are exceptionally small, demonstrating that the projection of the plane-wave electronic states onto the localized atomic-orbital basis set is highly accurate over the entire pressure range investigated [**133**].

**Table 6**. Mulliken atomic populations and effective valence charge (EVC) of $Nb_5Ir_3N$ at various pressures.

| *P* | Species | Ion | Mulliken atomic population | | | | Mulliken | Formal | EVC |
|---|---|---|---|---|---|---|---|---|---|
| | | | *s* | *P* | *d* | Total | charge | charge | (Mulliken) |
| 00 | N | 1 | 1.67 | 3.99 | 0 | 5.65 | -0.65 | -3 | 2.35 |
| | N | 2 | 1.67 | 3.99 | 0 | 5.65 | -0.65 | -3 | 2.35 |
| | Nb | 1 | 2.32 | 6.51 | 3.98 | 12.81 | 0.19 | +4 | 3.81 |
| | Nb | 2 | 2.32 | 6.51 | 3.98 | 12.81 | 0.19 | +4 | 3.81 |
| | Nb | 3 | 2.32 | 6.51 | 3.98 | 12.81 | 0.19 | +4 | 3.81 |
| | Nb | 4 | 2.32 | 6.51 | 3.98 | 12.81 | 0.19 | +4 | 3.81 |
| | Nb | 5 | 2.31 | 6.65 | 3.74 | 12.7 | 0.30 | +4 | 3.70 |
| | Nb | 6 | 2.31 | 6.65 | 3.74 | 12.7 | 0.30 | +4 | 3.70 |
| | Nb | 7 | 2.31 | 6.65 | 3.74 | 12.7 | 0.30 | +4 | 3.70 |
| | Nb | 8 | 2.31 | 6.65 | 3.74 | 12.7 | 0.30 | +4 | 3.70 |
| | Nb | 9 | 2.31 | 6.65 | 3.74 | 12.7 | 0.30 | +4 | 3.70 |
| | Nb | 10 | 2.31 | 6.65 | 3.74 | 12.7 | 0.30 | +4 | 3.70 |
| | Ir | 1 | 0.64 | 0.91 | 7.65 | 9.21 | -0.21 | +3 | 2.79 |
| | Ir | 2 | 0.64 | 0.91 | 7.65 | 9.21 | -0.21 | +3 | 2.79 |
| | Ir | 3 | 0.64 | 0.91 | 7.65 | 9.21 | -0.21 | +3 | 2.79 |
| | Ir | 4 | 0.64 | 0.91 | 7.65 | 9.21 | -0.21 | +3 | 2.79 |
| | Ir | 5 | 0.64 | 0.91 | 7.65 | 9.21 | -0.21 | +3 | 2.79 |
| | Ir | 6 | 0.64 | 0.91 | 7.65 | 9.21 | -0.21 | +3 | 2.79 |
| 10 | N | 1 | 1.66 | 3.99 | 0 | 5.65 | -0.65 | -3 | 2.35 |
| | N | 2 | 1.66 | 3.99 | 0 | 5.65 | -0.65 | -3 | 2.35 |
| | Nb | 1 | 2.3 | 6.45 | 4.02 | 12.77 | 0.23 | +4 | 3.77 |
| | Nb | 2 | 2.3 | 6.45 | 4.02 | 12.77 | 0.23 | +4 | 3.77 |
| | Nb | 3 | 2.3 | 6.45 | 4.02 | 12.77 | 0.23 | +4 | 3.77 |
| | Nb | 4 | 2.3 | 6.45 | 4.02 | 12.77 | 0.23 | +4 | 3.77 |
| | Nb | 5 | 2.3 | 6.62 | 3.77 | 12.69 | 0.31 | +4 | 3.69 |
| | Nb | 6 | 2.3 | 6.62 | 3.77 | 12.69 | 0.31 | +4 | 3.69 |
| | Nb | 7 | 2.3 | 6.62 | 3.77 | 12.69 | 0.31 | +4 | 3.69 |
| | Nb | 8 | 2.3 | 6.62 | 3.77 | 12.69 | 0.31 | +4 | 3.69 |
| | Nb | 9 | 2.3 | 6.62 | 3.77 | 12.69 | 0.31 | +4 | 3.69 |
| | Nb | 10 | 2.3 | 6.62 | 3.77 | 12.69 | 0.31 | +4 | 3.69 |
| | Ir | 1 | 0.65 | 0.95 | 7.66 | 9.25 | -0.25 | +3 | 2.75 |
| | Ir | 2 | 0.65 | 0.95 | 7.66 | 9.25 | -0.25 | +3 | 2.75 |
| | Ir | 3 | 0.65 | 0.95 | 7.66 | 9.25 | -0.25 | +3 | 2.75 |
| | Ir | 4 | 0.65 | 0.95 | 7.66 | 9.25 | -0.25 | +3 | 2.75 |
| | Ir | 5 | 0.65 | 0.95 | 7.66 | 9.25 | -0.25 | +3 | 2.75 |
| | Ir | 6 | 0.65 | 0.95 | 7.66 | 9.25 | -0.25 | +3 | 2.75 |
| 20 | N | 1 | 1.66 | 3.99 | 0 | 5.65 | -0.65 | -3 | 2.35 |
| | N | 2 | 1.66 | 3.99 | 0 | 5.65 | -0.65 | -3 | 2.35 |
| | Nb | 1 | 2.28 | 6.39 | 4.05 | 12.72 | 0.28 | +4 | 3.72 |
| | Nb | 2 | 2.28 | 6.39 | 4.05 | 12.72 | 0.28 | +4 | 3.72 |
| | Nb | 3 | 2.28 | 6.39 | 4.05 | 12.72 | 0.28 | +4 | 3.72 |
| | Nb | 4 | 2.28 | 6.39 | 4.05 | 12.72 | 0.28 | +4 | 3.72 |
| | Nb | 5 | 2.29 | 6.59 | 3.79 | 12.67 | 0.33 | +4 | 3.67 |
| | Nb | 6 | 2.29 | 6.59 | 3.79 | 12.67 | 0.33 | +4 | 3.67 |
| | Nb | 7 | 2.29 | 6.59 | 3.79 | 12.67 | 0.33 | +4 | 3.67 |
| | Nb | 8 | 2.29 | 6.59 | 3.79 | 12.67 | 0.33 | +4 | 3.67 |
| | Nb | 9 | 2.29 | 6.59 | 3.79 | 12.67 | 0.33 | +4 | 3.67 |
| | Nb | 10 | 2.29 | 6.59 | 3.79 | 12.67 | 0.33 | +4 | 3.67 |
| | Ir | 1 | 0.65 | 0.99 | 7.66 | 9.3 | -0.30 | +3 | 2.70 |
| | Ir | 2 | 0.65 | 0.99 | 7.66 | 9.3 | -0.30 | +3 | 2.70 |
| | Ir | 3 | 0.65 | 0.99 | 7.66 | 9.3 | -0.30 | +3 | 2.70 |
| | Ir | 4 | 0.65 | 0.99 | 7.66 | 9.3 | -0.30 | +3 | 2.70 |
| | Ir | 5 | 0.65 | 0.99 | 7.66 | 9.3 | -0.30 | +3 | 2.70 |
| | Ir | 6 | 0.65 | 0.99 | 7.66 | 9.3 | -0.30 | +3 | 2.70 |

As seen in **Table 6**, the calculated Mulliken charges reveal that the Nb atoms possess positive charges, whereas both N and Ir exhibit negative charges, demonstrating that Nb primarily acts as an electron-donating species while N and Ir serve as electron acceptors. At ambient pressure, the Mulliken charges of the two inequivalent Nb sites are approximately $+0.19e$ and $+0.30e$, respectively, while those of N and Ir are $-0.65e$ and $-0.21e$. The considerable difference between these effective atomic charges and their corresponding formal ionic charges indicates that the bonding cannot be regarded as purely ionic and instead possesses a substantial covalent contribution. With increasing pressure from 0 to 20 GPa, the Mulliken charges of Nb (1–4) increase from $+0.19e$ to $+0.28e$, while those of Nb (5–10) increase from $+0.30e$ to $+0.33e$. In contrast, the charge on Ir becomes progressively more negative, changing from $-0.21e$ to $-0.30e$, whereas the charge on N remains almost unchanged at $-0.65e$. These results indicate that the additional electronic charge released from Nb under compression is transferred predominantly toward the Ir atom.

The EVC further supports the mixed ionic–covalent nature of the bonding. The substantial non-zero values of EVC exhibit a pronounced deviation from the ideal ionic configuration and hence an appreciable covalent contribution. With increasing pressure, the EVC values of both Nb and Ir decrease slightly, whereas that of N remains nearly invariant, indicating that compression primarily modifies the Nb–Ir bonding environment while having a relatively weak influence on the local charge state of N.

### *3.4.3. Bond population and theoretical hardness*

The bond overlap population (BOP), when considered together with these Mulliken-charge and EVC results, can provide a more direct quantitative assessment of individual bond strengths. A zero or negative BOP value signifies negligible orbital overlap or antibonding interactions, indicating extremely weak or potentially unstable bonding. Under such conditions, the corresponding bond is generally unsuitable for a reliable estimation of the theoretical Vickers hardness [**134**]. In contrast, a larger positive BOP value indicates stronger orbital overlap and a greater degree of covalent character, whereas a smaller positive BOP value suggests a more pronounced ionic contribution to the chemical bond.

Based on the Mulliken BOP and bond hardness, the theoretical Vickers hardness ($H_V$) can be estimated by relating the bonding characteristics to the mechanical hardness of the material. The empirical model proposed by Gao *et al.* [**135**] establishes a correlation between bond properties and the Vickers hardness. The following equations are employed to calculate $H_V$:

$$\left.\begin{aligned} H_V &= \left[\prod \left(H_V^{\mu}\right)^{n^{\mu}}\right]^{1/\sum n^{\mu}} \\ H_V^{\mu} &= 740\, P^{\mu}\left(v_b^{\mu}\right)^{-5/3} \\ v_b^{\mu} &= (d^{\mu})^3 / \sum_{\mu}\left[(d^{\mu})^3 N_b^{\mu}\right] \end{aligned}\right\} \quad (22)$$

where $H_V^{\mu}$ is the hardness of the $\mu$-type bond, $n^{\mu}$ is the bond number of the $\mu$-type, $P^{\mu}$ is the Mulliken overlap population of the $\mu$-type bond, $v_b^{\mu}$ is the volume of the $\mu$-type bond, $d^{\mu}$ is the bond length, and $N_b^{\mu}$ is the total number of bonds in the unit cell per unit volume.

**Table 7**. Calculated Mulliken bond number $n^{\mu}$, bond length $d^{\mu}$, bond overlap population $P^{\mu}$, bond volume $v_b^{\mu}$ and bond hardness $H_v^{\mu}$ of $\mu$-type bond, total number of bond $N$, and Vickers hardness of $Nb_5Ir_3N$ electride.

| $P$ (GPa) | Bond | $n^{\mu}$ | $d^{\mu}$ (Å) | $P^{\mu}$ | $v_b^{\mu}$ (Å$^3$) | $H_v^{\mu}$ (GPa) | $N$ | $H_V$ (GPa) |
|---|---|---|---|---|---|---|---|---|
| 00 | N–Nb | 12 | 2.20055 | 0.26 | 2.3848 | 45.198 | 66 | 16.92 |
| | Nb–Ir (I) | 24 | 2.68193 | 0.19 | 4.3172 | 12.283 | | |
| | Nb–Ir (II) | 06 | 2.78621 | 0.41 | 4.8407 | 21.903 | | |
| | Nb–Ir (III) | 12 | 2.82644 | 0.18 | 5.0534 | 8.951 | | |
| | Nb–Ir (IV) | 06 | 2.96038 | 0.49 | 5.8064 | 19.331 | | |
| | Ir–Ir | 03 | 2.88543 | 0.56 | 5.3765 | 25.114 | | |
| | Nb–Nb | 02 | 2.56863 | -0.34 | - | - | | |
| | N–N | 01 | 2.56863 | -0.17 | - | - | | |
| 10 | N–Nb | 12 | 2.17984 | 0.26 | 2.3168 | 47.431 | 66 | 15.95 |
| | Nb–Ir (I) | 24 | 2.65620 | 0.15 | 4.1917 | 10.186 | | |
| | Nb–Ir (II) | 6 | 2.76874 | 0.43 | 4.7474 | 23.729 | | |
| | Nb–Ir (III) | 12 | 2.78155 | 0.15 | 4.8136 | 8.089 | | |
| | Nb–Ir (IV) | 6 | 2.91119 | 0.50 | 5.5185 | 21.470 | | |
| | Ir–Ir | 3 | 2.86536 | 0.60 | 5.2620 | 27.891 | | |
| | Nb–Nb | 2 | 2.53454 | -0.54 | - | - | | |
| | N–N | 1 | 2.53454 | -0.18 | - | - | | |
| 20 | N–Nb | 12 | 2.16201 | 0.26 | 2.2593 | 49.461 | 66 | 14.83 |
| | Nb–Ir (I) | 24 | 2.63343 | 0.12 | 4.0828 | 8.514 | | |
| | Nb–Ir (II) | 12 | 2.74551 | 0.12 | 4.6266 | 6.913 | | |
| | Nb–Ir (III) | 06 | 2.75054 | 0.44 | 4.6521 | 25.115 | | |
| | Nb–Ir (IV) | 06 | 2.87147 | 0.50 | 5.2930 | 23.016 | | |
| | Ir–Ir | 03 | 2.84616 | 0.65 | 5.1543 | 31.275 | | |
| | Nb–Nb | 02 | 2.50618 | -0.74 | - | - | | |
| | N–N | 01 | 2.50618 | -0.19 | - | - | | |

Out of the 66 bonds, 63 are used to calculate the Vickers hardness ($H_V$) at all pressures. The remaining three bonds are excluded from the hardness calculation due to negative populations. The calculated $H_V$ decreases monotonically from 16.92 GPa at ambient pressure to 15.95 GPa at 10 GPa and further to 14.83 GPa at 20 GPa (**Table 7**), indicating pressure-induced mechanical softening. Although hardness commonly increases with pressure due to enhanced atomic packing and stronger interatomic interactions, the present trend suggests that pressure modifies the bonding characteristics in a manner that facilitates plastic deformation. This behavior may arise from pressure-induced bond weakening, electron delocalization, and increased metallicity resulting in reduced resistance to indentation despite lattice compression [**86**].

### *3.4.4. Charge density difference (CDD)*

The interatomic bonding nature in a material can be explored from the electron density difference (EDD) around the different atoms in the compound, which displays the charge accumulation and depletion (electron loss) around each element. The pressure-dependent charge density difference (CDD) distributions of $Nb_5Ir_3N$ at 0, 10, and 20 GPa are presented in **Fig. 14**. The color scale for each map exhibits the electron density difference. The yellow/red and green color in CDD maps indicates high and low electron density, respectively. At ambient pressure, considerable charge redistribution is observed around the Nb, Ir, and N atoms, demonstrating substantial electronic interactions within the Nb–Ir–N framework (**Fig. 14a**).

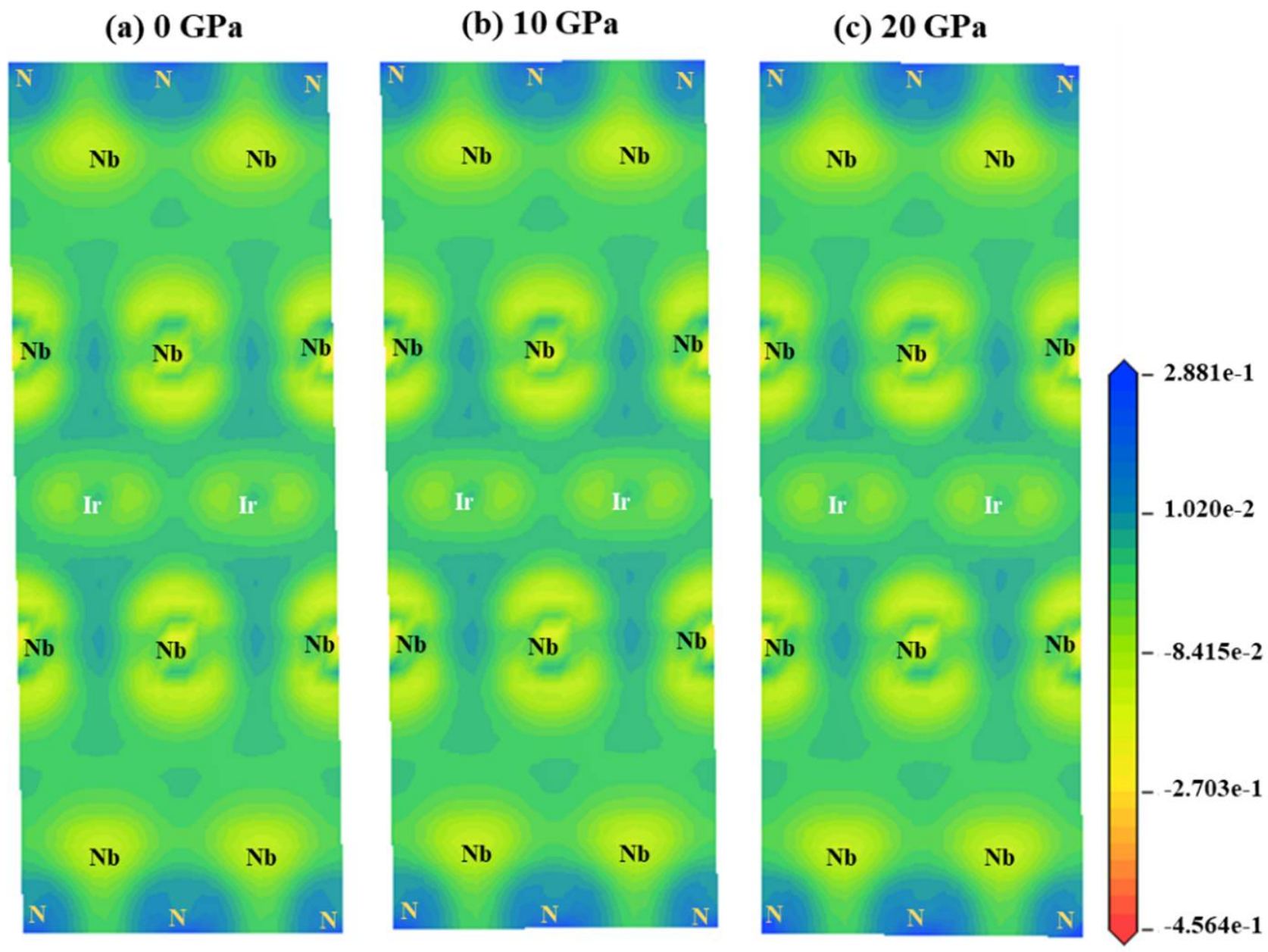

**Fig. 14**. Electron density distribution mapping of $Nb_5Ir_3N$ at various pressures.

Upon increasing pressure, the overall CDD topology remains essentially unchanged, although subtle modifications in the spatial distribution and magnitude of charge redistribution are evident, particularly around the Nb framework. Minimum EDD around N atoms for all pressures illustrates that charges are depleted around element. On the other hand, the yellow (approaches to red) color of Nb and Ir atoms signify electrons are donated by them. It is evident that the ionic bonding appears in Nb-N, polar-covalent character arising from the strong interaction between Nb 4*d* and N-2*p* states and Ir-N bonds. In contrary, Nb–Ir interactions are comparatively delocalized and can be associated with predominantly metallic bonding accompanied by covalent *d*-orbital hybridization.

### *3.5. Optical properties*

The optical response of materials provides essential information about their interaction with electromagnetic radiation and is significantly influenced by their electronic characteristics. Within the DFT framework, optical properties can be systematically evaluated from the calculated electronic band structure and momentum matrix elements between occupied and unoccupied electronic states. All calculations were carried out for incident photon energies ranging from 0 to 30 eV in <100> and <001> polarization directions, covering the infrared, visible, and ultraviolet (UV) spectral regions. Utilizing the metallic nature of $Nb_5Ir_3N$, a semi-empirical Drude model is adopted with a Gaussian smearing of 0.5 eV to precisely calculate the frequency-dependent dielectric constant. For both directions, a Drude damping of 0.05 eV and a plasma frequency of 10.0 eV is used.

The frequency-dependent dielectric function, which is the fundamental quantity describing the optical response, is generally expressed within the real $\varepsilon_1(\omega)$ and the imaginary part $\varepsilon_2(\omega)$ as expressed in **Eqs. 4-6**. The complex dielectric function $\varepsilon(\omega)$ provides insight into a material's linear optical response to electromagnetic radiation. The calculated $\varepsilon_1(\omega)$ and $\varepsilon_2(\omega)$ is depicted in **Fig. 15a**, exhibiting distinct and significant optical anisotropic features.

The pronounced negative value of $\varepsilon_1(\omega)$ at low photon energies for $Nb_5Ir_3N$, approaching divergence in the limit of $\omega \rightarrow 0$, is characteristic of metallic behavior and is consistent with the electronic structure calculations. The negative dielectric response originates from free-carrier contributions described by the Drude model. Since $\varepsilon_1(\omega)$ represents the dispersive contribution to the dielectric response, it provides information about the electronic polarizability of the system. Furthermore, the zero-crossing of $\varepsilon_1(\omega)$ near 27 eV in the UV region indicates a plasma-frequency-related transition, beyond which the material exhibits reduced optical absorption and enhanced transparency. In contrast, a pronounced $\varepsilon_2(\omega)$ response in the low-energy region is characteristic of metallic systems due to the contribution of free carriers. The electromagnetic energy dissipation capability of a material is enhanced when the absorptive component, $\varepsilon_2(\omega)$, is large, accompanied by significant changes in the dispersive component, $\varepsilon_1(\omega)$. Both intraband and interband electronic transitions contribute to $\varepsilon_2(\omega)$; however, intraband transitions dominate the low-energy optical response of metallic systems, particularly within the 0–2 eV energy range, whereas interband transitions govern the higher-energy features [**136**]. Furthermore, both $\varepsilon_1(\omega)$ and $\varepsilon_2(\omega)$ approach zero near 27 eV, corresponding to the plasma frequency region where several optical parameters exhibit critical changes. At this energy, the absorption coefficient (**Fig. 15c**) and reflectivity (**Fig. 15e**) decrease sharply, while the energy-loss function (**Fig. 15f**) exhibits a prominent peak, indicating the occurrence of plasma oscillation, which is characteristic feature for metals [**137**].

The calculated $\varepsilon_1(\omega)$ and $\varepsilon_2(\omega)$ are further utilized to derive the subsequent optical spectra, such as the refractive index $n(\omega)$, extinction coefficient $k(\omega)$, absorption coefficient $\alpha(\omega)$, conductivity $\sigma(\omega)$, reflectivity $R(\omega)$, and energy-loss function $L(\omega)$, which are represented in **Fig. 15b-f**. The refractive index $n(\omega)$ is a key optical parameter to describe how light propagates through a material and indicates the extent to which it slows down. As in **Fig. 15b**, the variation of $n(\omega)$ and $k(\omega)$ with photon energy of $Nb_5Ir_3N$ demonstrates a similar decreasing trend for all applied pressures. A steep decrease in both the $n(\omega)$ and $k(\omega)$ at low energies characterizes $Nb_5Ir_3N$ as typical metallic system for all studied pressures [**136**]. Both parameters reach their maximum values at zero photon energy, followed by a continuous reduction with increasing photon energy. This trend originates from the decreased contribution of electronic states available for optical excitation away from the Fermi level. The $k(\omega)$ provides information about the absorption capability of the material; therefore, larger $k(\omega)$ values indicate stronger attenuation of incident light. Moreover, the high $n(\omega)$ of $Nb_5Ir_3N$ in the visible region suggests favorable characteristics for optical coatings and optoelectronic display technologies.

The absorption coefficient $\alpha(\omega)$ exhibits a non-zero low-energy response along both electric field polarization directions (cf. **Fig. 15c**), which is consistent with the metallic electronic structure in $Nb_5Ir_3N$. The enhanced absorption for the <001> direction compared to the <100> direction indicates anisotropic optical behavior. Under applied pressure, the absorption peaks gradually shift toward higher photon energies (**Fig. 15c**), indicating pressure-induced modification of the electronic transition energies. A sharp decline in absorption occurs within the 26.5–28.0 eV energy range for pressures between 0 and 20 GPa, in agreement with the behavior of the dielectric functions and the plasma-response region. The apex peak of $\alpha(\omega)$ is observed at photon energy 7.78 eV as ~ $3.57\times10^5$ $cm^{-1}$ at 20 GPa for <001> polarization, along with significant absorption throughout the UV region. Such high value of $\alpha(\omega)$ in the order of $10^5$ $cm^{-1}$ indicate efficient photon absorption [**138**] and reflects strong interaction with incident radiation, suggesting promising optical characteristics for potential optoelectronic applications [**139**].

The real part of the optical conductivity $\sigma(\omega)$, shown in **Fig. 15d**, provides further insight into the electronic response of $Nb_5Ir_3N$ under photon excitation. The non-zero conductivity at zero photon energy confirms the metallic character of $Nb_5Ir_3N$ consistent with the dielectric function and absorption spectra. Along the <100> (<001>) polarization, the optical conductivity reaches values of ~ 8.9 (9.2) $\Omega^{-1}$ $cm^{-1}$ at 0 GPa and increases to about 9.3 (9.7) $\Omega^{-1}cm^{-1}$ at 20 GPa. It demonstrates that pressure can be used as an efficient tunning parameter of optical conductivity.

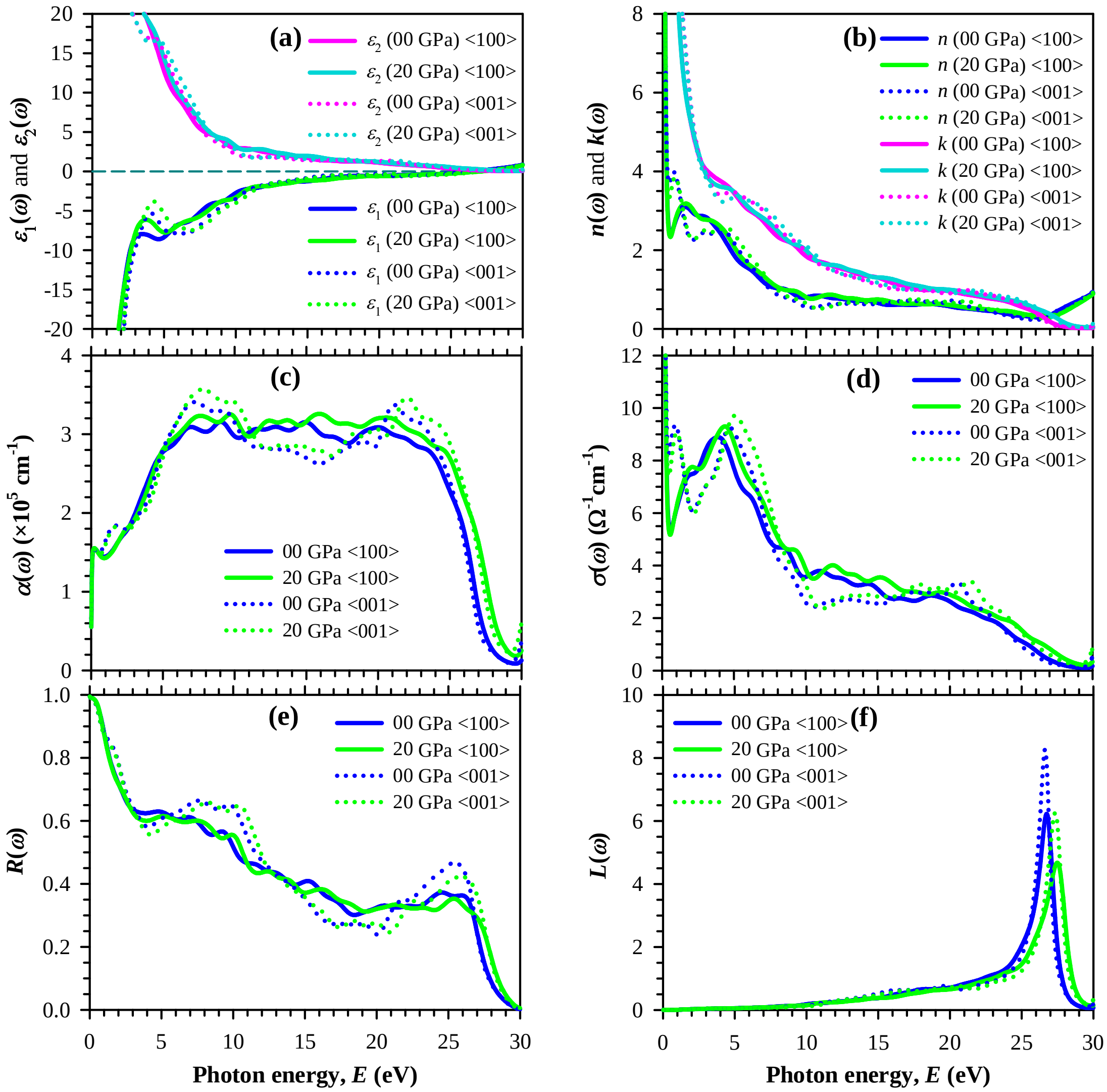

**Fig. 15**. Calculated (a) real ($\varepsilon_1$) and imaginary ($\varepsilon_2$) part of the dielectric function, (b) refractive index ($n$) and extinction coefficient ($k$), (c) absorption coefficient ($\alpha$), (d) conductivity ($\sigma$), (e) reflectivity, and (f) loss function of $Nb_5Ir_3N$ under pressure.

Reflectivity $R(\omega)$ spectra show enhanced values in the IR-region and maintain values above 57% across the visible spectrum (cf. **Fig. 15e**), suggesting strong reflective behavior compared with other reported metallic systems [140]. The reflectivity remains higher than approximately 50% up to 9.0 eV, indicating significant electromagnetic wave reflection over a wide UV-energy range. Therefore, $Nb_5Ir_3N$ may be considered a potential candidate for reflective optical coatings operating in IR, visible, and UV regions. Beyond 25.7 eV, the reflectivity decreases rapidly and approaches zero near 29 eV, coinciding with the energy region where the loss function reaches its maximum, indicating the onset of strong plasma excitation and reduced reflectance. Materials exhibiting relatively high reflectivity can reduce the absorption of incident electromagnetic radiation, thereby limiting surface heating and potentially improving thermal stability under irradiation environments.

The energy loss function $L(\omega)$ describes the energy dissipated by a fast electron while traversing a material. The $L(\omega)$ peak appears at $\varepsilon_2(\omega) < 1$ and $\varepsilon_1(\omega) = 0$ in the high energy region [101]. The prominent peaks observed in $L(\omega)$ correspond to plasma resonances arising from collective oscillations of electrons. Along the <100> direction, the maximum loss function peaks occur at approximately 26.7 eV and 27.5 eV for 0 and 20 GPa, respectively, whereas along the <001> direction, the corresponding peaks appear at ~ 26.5 eV and 27.3 eV (cf. **Fig. 15f**). The pressure-induced shift of these plasma peaks toward higher energies indicates modification of the electronic response due to compression. These peak positions coincide with the zero-crossing points of $\varepsilon_1(\omega)$, correspond to the trailing edges in $R(\omega)$ and $\sigma(\omega)$ spectra, confirming the occurrence of plasma resonance in $Nb_5Ir_3N$.

The coherent optical behavior of $Nb_5Ir_3N$ with electronic features, reported here for the first time, highlights its favorable characteristics and suggests its potential suitability for future optoelectronic applications.

### *3.6. Superconducting properties*

The emergence and strength of superconductivity are closely associated with the electronic structure, electron–phonon interactions, and many-body correlations of a material. In particular, the electron–phonon coupling constant ($\lambda_{ep}$), Coulomb pseudopotential ($\mu^*$), and density of states at the Fermi level [$N(E_F)$] are fundamental parameters that influence Cooper-pair formation. Their combined effects significantly govern the superconducting transition temperature ($T_c$) and the overall strength of superconductivity.

The effective Coulomb interaction is explained by the dimensionless Coulomb pseudopotential $\mu^*$, and can be calculated using the Bennemann-Garland formula [141]:

$$\mu^* = \frac{0.26\, N(E_F)}{1 + N(E_F)} \tag{23}$$

The electron–phonon coupling constant ($\lambda_{ep}$) can be estimated using the familiar McMillan-derived equation [142] from the reported value of $T_c$ as:

$$\lambda_{ep} = \frac{1.04 + \mu^* \, ln(\Theta_D/1.45T_c)}{(1 - 0.62\mu^*)\, ln(\Theta_D/1.45T_c) - 1.04} \tag{24}$$

**Table 8**. Calculated TDOS at the Fermi level [($N(E_F)$ in states/eV/f.u.], Debye temperatures ($\Theta_D$ in K), repulsive Coulomb pseudopotential ($\mu^*$), and electron-phonon coupling constant ($\lambda_{ep}$) of $Nb_5Ir_3N$ under pressure together with the available value.

| $P$ (GPa) | $N(E_F)$ | $\Theta_D$ | $\mu^*$ | $\lambda_{ep}$ | Ref. |
|---|---|---|---|---|---|
| 0 | 5.71 | 327.0 | 0.13 | 0.75 | [18][a] |
| 0 | 7.80 | 309.0 | 0.230 | -- | [This work] |
| 10 | 6.78 | 328.0 | 0.227 | -- | [This work] |
| 20 | 6.40 | 342.7 | 0.225 | -- | [This work] |

[a] Expt.

The estimation of $\lambda_{ep}$ is carried out from inverted McMillan-derived formula as **Eqn. 24**. In this approach, the experimentally observed $T_c$ = 8.7 K from [**18**] is utilized to estimate $\lambda_{ep}$ in the ground state. The calculated values of $\Theta_D$, $N(E_F)$, and $\mu^*$are listed in **Table 8**. The electron-phonon coupling constant can be expressed as: $\lambda_{ep} = N(E_F)V_{e-ph}$ [**143**]. As the phonon spectrum of $Nb_5Ir_3N$ shows a weak pressure variation, it is reasonable to assume that the strength of $V_{e-ph}$ does not vary significantly with pressure. With this assumption in mind, we suggest a decreasing trend in $\lambda_{ep}$ with pressure for $Nb_5Ir_3N$. The repulsive Coulomb pseudopotential $\mu^*$ is high for $Nb_5Ir_3N$ [**144**]. This parameter hinders Cooper pair formation. The variation of $\mu^*$ with pressure is rather weak. The Debye temperature which is positively correlated with superconducting $T_c$ also shows a weak pressure dependence. The largest impact of pressure on superconducting $T_c$ of $Nb_5Ir_3N$ is predicted due to the reduction of $N(E_F)$ with increasing pressure. Thus, we predict a weak pressure dependent decrement in $T_c$ for $Nb_5Ir_3N$. For conclusive insights on superconducting transition temperature under pressure, a detailed pressure dependent Eliashberg treatment via the electron-phonon spectral function is required, which lies beyond the scope of this paper.

## 4. Conclusions

This study comprehensively investigates the pressure-dependent physical properties of the ternary nitride superconductor $Nb_5Ir_3N$ over the 0–20 GPa range for the first time using first-principles calculations implemented in the CASTEP and Quantum ESPRESSO (QE) codes. The hexagonal phase of $Nb_5Ir_3N$ remains structurally and thermodynamically stable over the investigated pressure range, supported by negative formation energies, positive cohesive energies, and well-converged Birch–Murnaghan equations of state. Its mechanical and dynamical stability is further confirmed by the calculated elastic constants and the absence of negative phonon frequencies. Thermodynamic properties are calculated up to a temperature of 1000 K and found a consistent relationship with temperature, guiding further explorations into material behavior under varying thermal conditions. The evolution of the single-crystal elastic constants and polycrystalline elastic moduli under pressure provides detailed insight into the mechanical and anisotropic behavior of $Nb_5Ir_3N$. The calculated mechanical indices confirm its ductile nature, whereas the elastic anisotropy factors reveal that the compound remains elastically anisotropic throughout the investigated pressure range, with the anisotropy steadily decreasing under compression. Elasto-acoustic and hardness of $Nb_5Ir_3N$ are also investigated thoroughly under pressure. Minimum and lattice thermal conductivities, Debye and melting temperatures, and sound velocities all rises with pressure. The combination of high level of ductility, excellent machinability, dry-lubricity, substantial hardness together with very high melting temperature and small minimum phonon thermal conductivity make $Nb_5Ir_3N$ an intriguing system for structural and high-temperature thermal applications. The electronic band structure and density of states calculated both without and with spin–orbit coupling, confirm the metallic nature of $Nb_5Ir_3N$, while the SOC-induced energy splitting highlights the significant role of SOC on its electronic properties. For instance, the degeneracies on several high symmetry points (*e.g. G*, *K*, *L*, *H*) are lifted when SOC is considered. The calculated optical spectra indicate that there is substantial intraband contributions at low photon energies, while exhibiting strong UV absorption (~$10^5$ cm$^{-1}$) at higher photon energies. The combination of strong low-energy reflectivity and pronounced UV absorption suggests potential relevance for electromagnetic shielding and UV-related optical applications.

All the investigated physical properties of $Nb_5Ir_3N$ exhibit systematic variation with pressure, suggesting that pressure can be used as a useful tunning parameter to customize various properties of this compound for specific applications.

The insights obtained from this study are expected to encourage future experimental and theoretical investigations and assess the potential of $Nb_5Ir_3N$ as a promising material for multifunctional applications.

## Acknowledgements

M.A.H.S. acknowledges the fellowship from the Science and Technology Fellowship Trust, Ministry of Science and Technology, Bangladesh for his Ph.D. research. S.H.N. acknowledges the research grant (1151/5/52/RU/Science-07/19-20) from the Faculty of Science, University of Rajshahi, Bangladesh, which partly supported this work.

## Data availability

Data will be made available from the corresponding author on reasonable request.

## Declaration of Generative AI and AI-assisted technologies

In preparing this manuscript, the authors employed ChatGPT (OpenAI) to enhance clarity of the language and overall readability. Following this, the authors meticulously reviewed and refined the content, asserting full responsibility for the published article's final version.

## Declaration of competing interest

The authors declare that they have no known competing financial interests or personal relationships that could have appeared to influence the work reported in this paper.

## CRediT authorship contribution statement

**M.A.H. Shah**: Conceptualization, Investigation, Software, Methodology, Data curation, Visualization, Resources, Formal analysis, Writing–original draft; **J.H. Abir**: Software, Investigation, Formal analysis; **S.H. Naqib**: Conceptualization, Validation, Supervision, Software, Project administration, Writing-review & editing.